\documentclass[aps,prd,twocolumn,superscriptaddress,nofootinbib]{revtex4-2}

\usepackage{supertabular} 
\usepackage{placeins}
\usepackage{epsfig}
\usepackage{graphicx}

\usepackage{float}
\usepackage{mathtools}
\usepackage{array}
\usepackage{dcolumn}

\usepackage{latexsym}
\usepackage{amsmath}
\usepackage{amssymb}
\usepackage{amsfonts}
\usepackage{bm}
\usepackage{physics}

\usepackage{color}
\definecolor{purple}{rgb}{0.5,0,0.5}
\definecolor{blue}{rgb}{0.0,0,0.9}
\definecolor{prdblue}{rgb}{0.133,0.118,0.498}
\usepackage[colorlinks=true, pdfstartview=FitV, linkcolor=prdblue, citecolor= prdblue, urlcolor=prdblue]{hyperref}

\def\tstrut{\vrule height3.25ex depth0pt width0pt} 

\begin{document}


\title{Can one-loop corrections to the one-gluon exchange potential adequately describe the charmed meson spectrum?}

\author{A.~Capelo-Astudillo}
\email{andre.capelo@epn.edu.ec}
\affiliation{Departamento de Física, Escuela Politécnica Nacional, Quito 170143, Ecuador}

\author{T.~Aguilar}
\affiliation{Departamento de Física, Escuela Politécnica Nacional, Quito 170143, Ecuador}

\author{M.~Conde-Correa}
\affiliation{Departamento de Física, Escuela Politécnica Nacional, Quito 170143, Ecuador}

\author{A.~Duenas-Vidal}
\email[]{alvaro.duenas@epn.edu.ec}
\affiliation{Departamento de Física, Escuela Politécnica Nacional, Quito 170143, Ecuador}

\author{P.~G.~Ortega}
\email[]{pgortega@usal.es}
\affiliation{Departamento de F\'isica Fundamental, Universidad de Salamanca, E-37008 Salamanca, Spain}
\affiliation{Instituto Universitario de F\'isica 
Fundamental y Matem\'aticas (IUFFyM), Universidad de Salamanca, E-37008 Salamanca, Spain}

\author{J.~Segovia}
\email[]{jsegovia@upo.es}
\affiliation{Departamento de Sistemas F\'isicos, Qu\'imicos y Naturales, Universidad Pablo de Olavide, E-41013 Sevilla, Spain}

\date{\today}

\begin{abstract}
We investigate the charmed meson spectrum using a constituent quark model (CQM) with one-loop corrections applied to the one-gluon exchange (OGE) potential. The study aims to understand if the modified version of our CQM sufficiently account for the charmed meson spectrum observed experimentally, without invoking exotic quark and gluon configurations such as hybrid mesons or tetraquarks. Within this model, charmed mesons' masses are computed, comparing theoretical predictions to experimental data. The results, within uncertainties, suggest that our theoretical framework generally reproduces mass splittings and level ordering observed for charmed mesons. Particularly, large discrepancies between theory and experiment found in $P$-wave states are, at least, significantly ameliorated by incorporating higher-order interaction terms. Therefore, the findings emphasize that while the traditional quark model is limited in fully describing charmed mesons, enhanced potential terms may bridge the gap with experimental observations. The study contributes a framework for predicting excited charmed meson states for future experimental validation.
\end{abstract}


\maketitle


\section{INTRODUCTION}
\label{sec:introduction}

A simple analysis about the properties of mesons containing a single heavy quark, $Q=c$ or $b$, can be carried out in the limit of $m_Q\to\infty$. In such a regime, the heavy quark acts as a static color source for the rest of the heavy-light meson, \emph{i.e.} its spin $s_Q$ is decoupled from the total angular momentum of the light antiquark, $j_q$, and they are separately conserved. As a result, heavy-light mesons are grouped into doublets, each associated with a specific value of $j_q$ and parity. The members of each doublet differ from the orientation of $s_Q$ with respect to $j_q$ and they are degenerate in the heavy quark symmetry (HQS) limit~\cite{Isgur:1991wq}, whose mass degeneracy is broken at order $1/m_Q$.

For $Q\bar{q}$ states, and following HQS, one can write $\vec{j}_q=\vec{s}_q+\vec{\ell}$, where $s_q$ is the light antiquark spin and $\ell$ is its orbital angular momentum relative to the static heavy quark. Therefore, the lowest-lying $Q\bar{q}$ mesons correspond to $\ell=0$ with $j_q^P=\frac{1}{2}^{-}$. This doublet comprises two $S$-wave states with spin-parity $J^{P}=(0^{-},1^{-})$, where $\vec{J}=\vec{j}_q+\vec{s}_Q$. For $\ell=1$, it could be either $j_q^P=\frac{1}{2}^{+}$ or $j_q^P=\frac{3}{2}^{+}$, and thus the two corresponding doublets are $J^{P}=(0^+,1^+)$ and $J^{P}=(1^+,2^+)$, respectively. The mesons with $\ell=2$ are collected either in the $j_q^P=\frac{3}{2}^{-}$ doublet, consisting of states with $J^{P}=(1^-,2^-)$, or in the $j_q^P=\frac{5}{2}^{-}$ with $J^{P}=(2^-,3^-)$; and so forth and so on.

If we now focus on the spectrum of charmed mesons, $(c\bar n)$-states with $n=u$ or $d$ quark, it contains a number of long known and well established states collected in the Review of Particle Physics (RPP) of Particle Data Group (PDG)~\cite{ParticleDataGroup:2024cfk}. We find the lowest-lying $S$-wave states with quantum numbers $J^P=0^-$ and $1^-$, denoted as $D$ and $D^\ast$ mesons. The $P$-wave ground states with spin-parity quantum numbers $0^+$ ($D_0^\ast(2300)$), $1^+$ ($D_1(2420)$ and $D_1(2430)$) and $2^{+}$ ($D_2^\ast(2460)$) are also given in Ref.~\cite{ParticleDataGroup:2024cfk}. In addition, the PDG lists as well-established state, a highly-excited charmed meson, with spin-parity $J^P=3^-$, denoted as $D_3^{\ast}(2750)$. It was observed as a resonant substructure in the $B^0\to \bar{D}^0\pi^+\pi^-$ and $B^-\to D^+\pi^-\pi^-$ decays analyzed with the Dalitz plot technique~\cite{LHCb:2015klp, LHCb:2016lxy}. 

Over the past $15$ years or so, several new signals in the charmed meson sector have been observed. The now named $D_0(2550)$, $D_1^\ast(2600)$, $D_2(2740)$ and $D_3^\ast(2750)$ were observed for the first time by the BaBar collaboration in 2010~\cite{BaBar:2010zpy}, and were confirmed by the LHCb experiment with slightly different masses in 2013~\cite{LHCb:2013jjb}. Furthermore, the LHCb collaboration reported in Ref.~\cite{LHCb:2013jjb} two new higher $D$-meson excitations, $D_J^\ast(3000)$ and $D_J(3000)$, with natural and unnatural parities,\footnote{Natural parity means that the bosonic meson field behaves under reflection as $+1$ for even spin and $-1$ for odd spin; note then that, for heavy-light mesons, the superindex ``$\ast$" is used for those having natural parity.} respectively, collectively named by the PDG as $D(3000)^0$. In 2015, a new state $D_1^\ast(2760)$, with spin-parity quantum numbers $J^P=1^-$, was observed by the LHCb collaboration in the $D^+\pi^-$ channel by analyzing the $B^-\to D^+K^-\pi^-$ decay~\cite{LHCb:2015eqv}. Finally, there have been observed two more states which are not collected in the RPP of PDG. The first one is the named $D^\ast(2640)^\pm$ seen in $Z$ decays by  Abreu \emph{et al.}~\cite{DELPHI:1998oyl} but missing in the searches performed in Refs.~\cite{OPAL:2001xfs, ZEUS:2008nzg}, thus requiring confirmation. The second was observed in 2016 by the LHCb collaboration in the $D^+\pi^-$ channel when analyzing the $B^- \to D^+ \pi^- \pi^-$ decay~\cite{LHCb:2016lxy}; they assigned to this signal the name $D_2^\ast(3000)$ with spin-parity $J^P=2^+$ because its resonance parameters were inconsistent with the previously observed $D(3000)^0$~\cite{LHCb:2013jjb}. 

Theoretical predictions of the spectrum of charmed mesons dates from the early days of phenomenological quark models~\cite{Godfrey:1985xj, Zeng:1994vj, Gupta:1994mw, Lahde:1999ih}. In the last years, many studies have been carried out within different theoretical approaches such as lattice-regularized QCD~\cite{Mohler:2011ke, Moir:2013ub, Kalinowski:2015bwa, Cichy:2016bci, Gayer:2021xzv}, unitarized coupled-channels $T$-matrix analyses~\cite{Albaladejo:2016lbb, Du:2017zvv, Du:2020pui}, heavy meson effective theory~\cite{Wang:2010ydc, Wang:2013tka}, Regge-based phenomenology~\cite{Li:2007px, Chen:2018nnr} and phenomenological quark models~\cite{DiPierro:2001dwf, Godfrey:2005ww, Close:2005se, Close:2006gr, Zhang:2006yj, Wei:2006wa, Ebert:2009ua, Song:2015fha, Ferretti:2015rsa, Liu:2016efm}. This is mainly because two reasons; the first one is the recent experimental measurements in the subject which provide sustained progress in the field as well as the breadth and depth necessary for a vibrant theoretical research environment. The second is related mostly with the fact that $D_0^\ast(2300)$ and $D_1(2420)$ charmed mesons, which belong to the doublet $j_{q}^{P}=\frac{1}{2}^{+}$ predicted by HQS, have surprisingly light masses, compared with naive quark model expectations, and are located below $D\pi$ and $D^{\ast}\pi$ thresholds, respectively. This implies that these states are narrow. These facts have stimulated a fruitful line of research, suggesting that their structure is much richer than what one might guess assuming the $Q\bar{q}$ picture~\cite{Barnes:2003dj, vanBeveren:2003kd, Vijande:2006hj}.

The quark model has been notably successful in describing the heavy quark-antiquark system since the early days of charmonium studies (see, for example, Refs.~\cite{Eichten:1978tg, Eichten:1979ms, Gupta:1982kp, Gupta:1983we, Gupta:1984jb, Gupta:1984um, Kwong:1987ak, Kwong:1988ae}). Moreover, predictions from this framework on the properties of heavy quarkonia, including those related to decays and interactions, have proven highly valuable for guiding experimental searches. Additionally, the quark model's adaptability makes it well-suited for exploratory research on exotic matter. Thus, the theoretical results presented here are based on a constituent quark model (CQM), initially proposed in Ref.~\cite{Vijande:2004he}, and recently applied to conventional mesons containing heavy quarks, capturing a broad range of physical observables related to spectra~\cite{Segovia:2008zz, Segovia:2016xqb, Segovia:2015dia, Ortega:2020uvc}, strong decays~\cite{Segovia:2012cd, Segovia:2009zz, Segovia:2013kg}, hadronic transitions~\cite{Segovia:2014mca, Segovia:2015raa, Martin-Gonzalez:2022qwd}, and both electromagnetic and weak reactions~\cite{Segovia:2011zza, Segovia:2011dg, Segovia:2012yh}. To improve the accuracy of mass splittings, we adopt the approach in Ref.~\cite{Lakhina:2006fy} and incorporate one-loop corrections to the One-Gluon Exchange (OGE) potential as derived by Gupta and Radford~\cite{Gupta:1981pd}. These corrections include, for the first time in the perturbative series, a spin-dependent term that impacts only on mesons made by quarks of different flavors. Our primary objective is to determine whether the entire spectrum of experimentally observed charmed mesons can be described within the quark-antiquark model alone, without needing to invoke more exotic configurations.

The manuscript is organized as follows. After this introduction, the theoretical framework is briefly presented in Sec~\ref{sec:theory}. Section~\ref{sec:results} is mainly devoted to the analysis and discussion of our results. Finally, we summarize and draw some conclusions in Sec.~\ref{sec:summary}.


\section{THEORETICAL FORMALISM}
\label{sec:theory} 

The dynamical braking of chiral symmetry in Quantum Chromodynamics (QCD) is responsible, among other phenomena, of generating constituent quark masses and Goldstone-boson exchanges between light quarks. This together with the one-gluon exchange interaction and the color confining force consist on the main pieces of our constituent quark model~\cite{Vijande:2004he, Segovia:2013wma}.

Under chiral transformations, the following Lagrangian
\begin{equation}
{\mathcal L} = \bar{\psi}(i\, {\slash\!\!\! \partial} - M(q^{2})U^{\gamma_{5}})\,\psi \,,
\label{eq:ChiralL}
\end{equation}
is invariant~\cite{Diakonov:2002fq}. In Eq.~\eqref{eq:ChiralL}, $M(q^2)$ is the dynamical momentum-dependent constituent quark mass and $U^{\gamma_5} = e^{i\lambda _{a}\phi^{a}\gamma_{5}/f_{\pi}}$, with $\phi=\{\vec \pi,K,\eta_8\}$, is the matrix of Goldstone-boson fields that can be expanded as
\begin{equation}
U^{\gamma _{5}} = 1 + \frac{i}{f_{\pi}} \gamma^{5} \lambda^{a} \phi^{a} - \frac{1}{2f_{\pi}^{2}} \phi^{a} \phi^{a} + \ldots
\end{equation}
One can guess that the first term of the expansion provides the constituent quark mass, the second gives rise to one-boson exchange interactions between light quarks and the main contribution of the third term comes from the two-pion exchange which is simulated in our case by means of a scalar-meson exchange interaction. In the presence of heavy quarks, chiral symmetry is explicitly broken and Goldstone-boson exchanges do not appear. However, it constrains the model parameters through the light-meson phenomenology~\cite{Segovia:2008zza}.

At energy scales higher than that of dynamical breaking of chiral symmetry, the CQM incorporates QCD perturbative effects by taking into account one-gluon fluctuations around the instanton vacuum through the vertex Lagrangian
\begin{equation}
{\mathcal L}_{qqg} = i\sqrt{4\pi\alpha_{s}} \, \bar{\psi} \gamma_{\mu} G^{\mu}_{c} \lambda^{c} \psi \,,
\label{Lqqg}
\end{equation}
with $\lambda^{c}$ the $SU(3)$ color matrices and $G^{\mu}_{c}$ the gluon field. The $\alpha_{s}$ is a scale-dependent effective strong coupling constant that allows a comprehensive description of light, strange and heavy meson spectra~\cite{Vijande:2004he, Segovia:2013wma}:
\begin{equation}
\alpha_{s}(\mu_{ij})=\frac{\alpha_{0}}{\ln\left(\frac{\mu_{ij}^{2}+\mu_{0}^{2}}{\Lambda_{0}^{2}} \right)},
\end{equation}
in which $\mu_{ij}$ is the reduced mass of the meson's constituent $q\bar{q}$ pair and $\alpha_{0}$, $\mu_{0}$ and $\Lambda_{0}$ are parameters of the quark model.

The potential derived from Eq.~\eqref{Lqqg} contains central, tensor, and spin-orbit contributions given by
\begin{widetext}
\begin{align}
V_{\text{OGE}}^{\text{C}}(\vec{r}_{ij})= & \frac{1}{4}\alpha_{s}(\vec{\lambda}_{i}^{c}\cdot \vec{\lambda}_{j}^{c})\left[ \frac{1}{r_{ij}}-\frac{1}{6m_{i}m_{j}} (\vec{\sigma}_{i}\cdot\vec{\sigma}_{j}) \frac{e^{-r_{ij}/r_{0}(\mu_{ij})}}{r_{ij}r_{0}^{2}(\mu_{ij})}\right], \nonumber \\
V_{\text{OGE}}^{\text{T}}(\vec{r}_{ij})= & -\frac{1}{16}\frac{\alpha_{s}}{m_{i}m_{j}} (\vec{\lambda}_{i}^{c}\cdot\vec{\lambda}_{j}^{c})\left[\frac{1}{r_{ij}^{3}}-\frac{e^{-r_{ij}/r_{g}(\mu_{ij})}}{r_{ij}}\left(\frac{1}{r_{ij}^{2}}+\frac{1}{3r_{g}^{2}(\mu_{ij})}+\frac{1}{r_{ij}r_{g}(\mu_{ij})}\right)\right]S_{ij}, \nonumber \\
V_{\text{OGE}}^{\text{SO}}(\vec{r}_{ij}) = & -\frac{1}{16}\frac{\alpha_{s}}{m_{i}^{2}m_{j}^{2}}(\vec{\lambda}_{i}^{c}\cdot \vec{\lambda}_{j}^{c})\left[\frac{1}{r_{ij}^{3}}-\frac{e^{-r_{ij}/r_{g}(\mu_{ij})}}{ r_{ij}^{3}} \left(1+\frac{r_{ij}}{r_{g}(\mu_{ij})}\right)\right] \times \nonumber \\
&
\times \left[((m_{i}+m_{j})^{2}+2m_{i}m_{j})(\vec{S}_{+}\cdot\vec{L})+(m_{j}^{2} -m_{i}^{2}) (\vec{S}_{-}\cdot\vec{L}) \right],
\end{align}
\end{widetext}
where $S_{ij}=3(\vec{\sigma}_{i}\cdot\hat{r}_{ij})(\vec{\sigma}_{j}\cdot
\hat{r}_{ij})-\vec{\sigma}_{i}\cdot\vec{\sigma}_{j}$ is the quark tensor
operator and $\vec{S}_{\pm} = \frac{1}{2} (\vec{\sigma}_{i} \,\pm\, \vec{\sigma}_{j})$ are the symmetric and antisymmetric spin-orbit operators, respectively. Besides, $r_{0}(\mu_{ij})=\hat{r}_{0}\frac{\mu_{nn}}{\mu_{ij}}$ and $r_{g}(\mu_{ij})=\hat{r}_{g}\frac{\mu_{nn}}{\mu_{ij}}$ are regulators which depend on $\mu_{ij}$, which is again the reduced mass of the meson's constituent $q\bar{q}$ pair. The contact term of the central potential has been regularized as $\delta(\vec{r}_{ij}) \approx (1/4\pi r_{0}^{2}) \cdot e^{-r_{ij}/r_{0}}/r_{ij}$.

To improve the description of charmed mesons, we follow the proposal of Ref.~\cite{Lakhina:2006fy} and include one-loop corrections to the OGE potential as derived by Gupta {\it et al.}~\cite{Gupta:1981pd}. As in the case of $V_{\rm OGE}$, $V_{\rm OGE}^{\rm 1-loop}$ contains central, tensor and spin-orbit contributions, given by~\cite{Ortega:2016mms}
\begin{widetext}
\begin{equation}
\begin{split}
&
V_{\text{OGE}}^{\rm 1-loop,C}(\vec{r}_{ij})=0, \\
&
\begin{split}
V_{\text{OGE}}^{\rm 1-loop,T}(\vec{r}_{ij}) = \frac{C_{F}}{4\pi} \frac{\alpha_{s}^{2}}{m_{i}m_{j}}\frac{1}{r^{3}}S_{ij} 
& 
\left[\frac{b_{0}}{2}\left(\ln(\mu r_{ij})+\gamma_{E}-\frac{4}{3}\right)+\frac{5}{12}b_ {0}-\frac{2}{3}C_{A} \right. \\ 
& 
\left. +\frac{1}{2}\left(C_{A}+2C_{F}-2C_{A}\left(\ln(\sqrt{m_{i}m_{j}}\,r_{ij})+\gamma_{E}-\frac{4}{3}\right)\right)\right],
\end{split} \\
&
\begin{split}
&V_{\text{OGE}}^{\rm 1-loop,SO}(\vec{r}_{ij})=\frac{C_{F}}{4\pi} \frac{\alpha_{s}^{2}}{m_{i}^{2}m_{j}^{2}}\frac{1}{r^{3}}\times \\
&
\begin{split} 
\times\Bigg\lbrace (\vec{S}_{+}\cdot\vec{L}) & \Big[ \left((m_{i}+m_{j})^{2}+2m_{i}m_{j}\right)\left(C_{F}+C_{A}-C_{A} \left(\ln(\sqrt{m_{i}m_{j}}\,r_{ij})+\gamma_{E}\right)\right) \\
&
+4m_{i}m_{j}\left(\frac{b_{0}}{2}\left(\ln(\mu r_{ij})+\gamma_{E}\right)-\frac{1}{12}b_{0}-\frac{1}{2}C_{F}-\frac{7}{6}C_{A} +\frac{C_{A}}{2}\left(\ln(\sqrt{m_{i}m_{j}}\,r_{ij})+\gamma_{E}\right)\right)
\\
&
+\frac{1}{2}(m_{j}^{2}-m_{i}^{2})C_{A}\ln\left(\frac{m_{j}}{m_{i}}\right)\Big] 
\end{split} \\
&
\begin{split}
\,\,\,\,\,\,\,+(\vec{S}_{-}\cdot\vec{L}) & \Big[(m_{j}^{2}-m_{i}^{2})\left(C_{F}+C_{A}-C_{A}\left(\ln(\sqrt{m_{i}m_{j}}\,r_ {ij})+\gamma_{E}\right)\right) \\
&
+\frac{1}{2}(m_{i}+m_{j})^{2}C_{A}\ln\left(\frac{m_{j}}{m_{i}}\right)\Big] \Bigg\rbrace,
\end{split}
\end{split}
\end{split}
\label{eq:1loop}
\end{equation}
\end{widetext}
where $C_{F}=4/3$, $C_{A}=3$, $b_{0}=9$, $\gamma_{E}=0.5772$ and the scale $\mu\approx1\,{\rm GeV}$.

Finally, an important non-perturbative term of our CQM is color confining interaction between quarks and antiquarks to ensure colorless hadrons. The potential used here is linearly-rising for short interquark distances, but acquires a plateau at large distances to mimic the effect of sea quarks, which induces the breakdown of the color binding string~\cite{Bali:2005fu}. Its explicit expression is
\begin{widetext}
\begin{align}
V_{\rm CON}^{\rm C}(\vec{r}_{ij}) =& \left[ -a_{c}(1-e^{-\mu_{c}r_{ij}})+\Delta \right] (\vec{\lambda}_{i}^{c}\cdot\vec{\lambda}_{j}^{c}), \nonumber \\
V_{\rm CON}^{\rm SO}(\vec{r}_{ij})= & -\left(\vec{\lambda}_{i}^{c}\cdot\vec{\lambda}_{j}^{c} \right)  \frac{a_{c}\mu_{c}e^{-\mu_{c}r_{ij}}}{4m_{i}^{2}m_{j}^{2}r_{ij}}\left[((m_{i}^{2 }+m_{j}^{2})(1-2a_{s}) +4m_{i}m_{j}(1-a_{s}))(\vec{S}_{+}\cdot\vec{L})\right. \nonumber \\
&
\left. +(m_{j}^{2}-m_{i}^{2})(1-2a_{s})(\vec{S}_{-}\cdot\vec{L}) \right], 
\end{align}
\end{widetext}
where the model parameters are $a_{c}$, $\Delta$, $\mu_{c}$ and $a_{s}$, being the last one the mixture between scalar and vector Lorentz structures of the confinement.

Among the different methods to solve the Schr\"odinger equation in order to find the quark-antiquark bound states, we use the Gaussian Expansion Method~\cite{Hiyama:2003cu} because it provides enough accuracy and makes the subsequent evaluation of matrix elements easier. This procedure provides the radial wave function solution of the Schr\"odinger equation as an expansion in terms of basis functions
\begin{equation}
R_{\alpha}(r)=\sum_{n=1}^{n_{max}} c_{n}^\alpha \phi^G_{nl}(r),
\end{equation} 
where $\alpha$ refers to the channel quantum numbers. Following Ref.~\cite{Hiyama:2003cu}, we employ Gaussian trial functions with ranges in geometric progression. This enables the optimization of ranges employing a small number of free parameters. Moreover, the geometric progression is dense at short distances, so that it allows the description of the dynamics mediated by short range potentials. The fast damping of the gaussian tail is not a problem, since we can choose the maximal range much longer than the hadronic size. The coefficients, $c_{n}^\alpha$, and the eigenvalue, $E$, are determined from the Rayleigh-Ritz variational principle
\begin{equation}
\sum_{n=1}^{n_{max}} \left[\left(T_{n'n}^\alpha-EN_{n'n}^\alpha\right)
c_{n}^\alpha+\sum_{\alpha'}
\ V_{n'n}^{\alpha\alpha'}c_{n}^{\alpha'}=0\right],
\end{equation}
where $T_{n'n}^\alpha$, $N_{n'n}^\alpha$ and $V_{n'n}^{\alpha\alpha'}$ are the matrix elements of the kinetic energy, the normalization and the potential,  respectively. The matrices $T_{n'n}^\alpha$ and $N_{n'n}^\alpha$ are diagonal whereas the mixing between different channels is given by $V_{n'n}^{\alpha\alpha'}$.

Model estimates of the mean momentum, $\langle p \rangle$, of a light constituent quark, with mass $M$, inside a meson typically yield $\langle p \rangle \sim M$. It might therefore be argued that bound-state calculations involving light quark systems should only be undertaken within models that, at some level, incorporate relativity. This potential weakness of the nonrelativistic quark model has long been considered. For example, Ref.~\cite{Capstick:1985xss} remarks that a non-relativistic treatment of quark motion is inaccurate. However, using scales that are internally consistent, it is not ultra-relativistic. Therefore, the non-relativistic approximation must be useful. The point is also canvassed in Ref.~\cite{Manohar:1983md}, which opens with the question ``Why does the non-relativistic quark model work?" and proceeds to provide a range of plausible answers. These discussions are complemented by Ref.~\cite{Lucha:1991vn}, which devotes itself to ``The significance of the treatment of relativistically moving constituents by an effective non-relativistic Schr\"odinger equation [...]". The conclusion of these discourses and many others is simple: the non-relativistic model has proved very useful, unifying a wide range of observables within a single framework.

This last observation provides our rationale for employing a non-relativistic model for the analysis herein. Namely, we take a pragmatic view: the non-relativistic quark model is a useful tool. The practical reason for its success is simple: the model has some parameters; they are fitted to a body of data; and, consequently, on this domain, the model cannot be wrong numerically. If one adds relativistic effects in one way or another, there are similar parameters in the new potential. They, too, are fitted to data; and hence the resulting model cannot produce results that are very different from the original non-relativistic version. The values of the parameters in the potential are modified, but the potential is not observable, so nothing substantive is altered.


\begin{table}[!t]
\caption{\label{tab:parameters} Constituent quark model parameters.}
\begin{ruledtabular}
\begin{tabular}{lrrr}
& & Original set & fine-tuned set \\[2ex]
Quark masses & $m_{n}$ (MeV) & $313$ & $313$ \\
             & $m_{c}$ (MeV) & $1763$ & $1763$ \\[2ex]
OGE & $\alpha_{0}$ & $2.118$ & $2.118$ \\
    & $\Lambda_{0}$ $(\mbox{fm}^{-1})$ & $0.113$ & $0.113$ \\
    & $\mu_{0}$ (MeV) & $36.976$ & $36.976$ \\
    & $\hat{r}_{0}$ (fm) & $0.181$ & $0.181$ \\
    & $\hat{r}_{g}$ (fm) & $0.259$ & $0.259$ \\[2ex] 
Confinement & $a_{c}$ (MeV) & $507.4$ & $478.0$ \\
	       & $\mu_{c}$ $(\mbox{fm}^{-1})$ & $0.576$ & $0.551$ \\
	       & $\Delta$ (MeV) & $184.432$ & $178.019$ \\
	       & $a_{s}$ & $0.81$ & $0.81$ \\
\end{tabular}
\end{ruledtabular}
\end{table}

\section{RESULTS}
\label{sec:results}

Model parameters relevant for this analysis are shown in Table~\ref{tab:parameters}. As stated in the Introduction, our main objective is to assess whether the full spectrum of experimentally observed charmed mesons can be roughly explained within the quark-antiquark model, without requiring more exotic configurations. All model parameters were constrained based on prior investigations of hadron phenomenology (see, for instance, Refs.~\cite{Vijande:2004he, Segovia:2013wma, Segovia:2016xqb}). However, for the lightest heavy-light meson sector, we must acknowledge that these are not the most suitable, and we have therefore taken the liberty of slightly modifying those that influence the slope of the linear confinement potential. This is why we show a column of finely tuned parameters in Table~\ref{tab:parameters}.

There are two types of theoretical uncertainties in our results: one is intrinsic to the numerical algorithm and the other is related to the way the model parameters are fixed. The numerical error is negligible and, as mentioned above, the model parameters are adjusted to reproduce a certain number of hadron observables within a determinate range of agreement with experiment. It is therefore difficult to assign an error to these parameters and consequently to the quantities calculated using them. To assess this, the results presented in this manuscript show a theoretical uncertainty of $(10-20)\%$ in the meson's mass.

\begin{table}[!t]
\caption{\label{tab:spectrum} Charmed meson masses, in MeV, from constituent quark model (CQM) and experiment~\cite{ParticleDataGroup:2024cfk, DELPHI:1998oyl, LHCb:2013jjb}. We show, for CQM's energy levels, the quark-antiquark value taking into account the one-gluon exchange potential ${\cal O}(\alpha_s)$ and including its one-loop correction ${\cal O}(\alpha_s^2)$. For experiment, we distinguish between well established states (\cite{ParticleDataGroup:2024cfk}) and those levels which still need confirmation and so have been omitted from the summary table (\cite{ParticleDataGroup:2024cfk}$^*$).}
\scalebox{0.95}{
\begin{tabular}{lcccccl}
\hline\hline
\tstrut
Meson & $J^P$ & n & The. ${\cal O}(\alpha_s)$ & The. ${\cal O}(\alpha_s^2)$ & Exp. & Ref. \\
\hline
\tstrut
$D$ & $0^-$ & 1 & $1868$ & $1868$ & $1867.95\pm0.27$ & \cite{ParticleDataGroup:2024cfk} \\
& & 2 & $2619$ & $2619$ & $2549\pm19$ & \cite{ParticleDataGroup:2024cfk}$^*$ \\
& & 3 & $3053$ & $3053$ & & \\[2ex]
$D_0^\ast$ & $0^+$ & 1 & $2445$ & $2281$ & $2343\pm10$ & \cite{ParticleDataGroup:2024cfk} \\ 
& & 2 & $2934$ & $2820$ & & \\
& & 3 & $3252$ & $3172$ & & \\[2ex]
$D^\ast$ & $1^-$ & 1 & $1982$ & $1977$ & $2009.12\pm0.04$ & \cite{ParticleDataGroup:2024cfk} \\ 
& & 2 & $2677$ & $2675$ & $2627\pm10$ & \cite{ParticleDataGroup:2024cfk}$^*$ \\
& & 3 & $2841$ & $2810$ & $2781\pm18\pm13$ & \cite{ParticleDataGroup:2024cfk}$^*$ \\[2ex]
$D_1$ & $1^+$ & 1 & $2410$ & $2438$ & $2422.1\pm0.8$ & \cite{ParticleDataGroup:2024cfk} \\ 
& & 2 & $2519$ & $2461$ & $2412\pm9$ & \cite{ParticleDataGroup:2024cfk} \\
& & 3 & $2918$ & $2940$ & & \\[2ex]
$D_2$ & $2^-$ & 1 & $2736$ & $2744$ & $2747\pm6$ & \cite{ParticleDataGroup:2024cfk}$^*$ \\ 
& & 2 & $2876$ & $2865$ & & \\
& & 3 & $3125$ & $3132$ & & \\[2ex]
$D_2^\ast$ & $2^+$ & 1 & $2452$ & $2483$ & $2461.1\pm0.7$ & \cite{ParticleDataGroup:2024cfk} \\ 
& & 2 & $2944$ & $2966$ & & \\
& & 3 & $3108$ & $3095$ & & \\[2ex]
$D_3^\ast$ & $3^-$ & 1 & $2767$ & $2783$ & $2763.1\pm3.2$ & \cite{ParticleDataGroup:2024cfk} \\ 
& & 2 & $3145$ & $3157$ & & \\
& & 3 & $3309$ & $3303$ & & \\[2ex]
$D_3$ & $3^+$ & 1 & $2995$ & $2998$ & & \\ 
& & 2 & $3130$ & $3127$ & & \\
& & 3 & $3296$ & $3299$ & & \\
\hline
\tstrut
$D^\ast(2640)^\pm$ & $?^?$ & & & & $2637\pm2\pm6$ & \cite{DELPHI:1998oyl} \\
$D(3000)^0$ & $?^?$ & & & & $3214\pm29\pm49$ & \cite{LHCb:2013jjb} \\
\hline\hline
\end{tabular}}
\end{table}

In Table~\ref{tab:spectrum}, we show the charmed meson masses, in MeV, from constituent quark model (CQM) and experiment~\cite{ParticleDataGroup:2024cfk, DELPHI:1998oyl, LHCb:2013jjb}. We show, for CQM's energy levels, the quark-antiquark value taking into account the one-gluon exchange potential ${\cal O}(\alpha_s)$ and including its one-loop correction ${\cal O}(\alpha_s^2)$. For experiment, we distinguish between well established states (\cite{ParticleDataGroup:2024cfk}) and those levels which still need confirmation and so have been omitted from the summary table (\cite{ParticleDataGroup:2024cfk}$^*$).

Two charmed mesons with quantum numbers $J^P=0^-$ have been experimentally observed, $D$ and $D_0(2550)$. The first one is the ground level of charmed mesons and it is well established in the RPP of PDG~\cite{ParticleDataGroup:2024cfk}. The second is still omitted from the summary table because even though two experiments observed this state its mass is different. Our theoretical prediction is slightly higher than the average mass reported by the RPP of PDG~\cite{ParticleDataGroup:2024cfk}; note that the experimental masses measured until now go from $2518$ to $2580$~MeV for this state. Another important feature to highlight is that ${\cal O}(\alpha_s^2)$ OGE corrections are zero in this $J^P$-channel and thus our na\"ive quark model must predict correctly these two states from the global fit of hadron phenomenology.

The partner of the $D$-meson which belongs to the $j_q^P=\frac{1}{2}^{-}$ doublet in heavy quark spin symmetry is the $D^\ast$ meson. As one can see in Table~\ref{tab:spectrum}, there are three candidates: $D^\ast$, $D_1^\ast(2600)$ and $D_1^\ast(2760)$; the first one is well established in PDG the other two are omitted from the summary table since they need confirmation. The $1$-loop OGE corrections are small to moderate in this channel producing mass shifts from $2$ to $30$ MeV. One may state that our results for $J^P=1^-$ channel agrees reasonably well with the experimental masses reported until now.

There are four $P$-wave states measured experimentally and denoted in the RPP of PDG as $D_0^\ast(2300)$, $D_1(2420)$, $D_1(2430)$, $D_2^\ast(2460)$. As one can see in Table~\ref{tab:spectrum}, our theoretical results reproduce correctly the level ordering and they are also in global agreement with the experimental reported masses once the one-loop OGE corrections are incorporated. The addition of the ${\cal O}(\alpha_s^2)$ OGE corrections was proposed by Lakhina \emph{et al.} in~\cite{Lakhina:2006fy} motivated by the fact that in the one-loop computation there is a spin-dependent term which affects only to mesons with different flavor quarks and it is not negligible for $P$-wave states where theory and experiment find their most significant differences. We demonstrate herein that \emph{na\"ive} quark models cannot reproduce $P$-wave charmed meson spectrum but, instead of resorting first to more complicated solutions such as exotic hadron structures, one should investigate the possibility of having missed potential terms that may be relevant for this sector.	 

The RPP of PDG reports two more charmed mesons with well established spin-parity quantum numbers, the $D_2(2740)$ and $D_3^\ast(2750)$ mesons. The first one is omitted from the summary table whereas the second is a well established charmed meson. Theoretically, both are dominant $D$-wave states whose masses are close to the experimental measurements; therefore, we may confirm that the experimental assignment is plausible. When incorporating the ${\cal O}(\alpha_s^2)$ OGE corrections, the theoretical masses of these states grow moderately but the change is not dramatic.

Focusing now on the two states whose quantum numbers have not been assigned (see the bottom part of Table~\ref{tab:spectrum}). The $D^\ast(2640)^\pm$ seems to have a mass similar to the expected one for the first excitation of the $D^\ast$ meson. In fact, there is no other possible case attending to the mass only. The $D(3000)^0$, whose mass is actually $(3214\pm29\pm49)\,\text{MeV}$, could be fitted as the first excitation of either $D_3^\ast$ or $D_3$, but could be also assigned as the second excitation of either $D_2$ or $D_2^\ast$ mesons. 

\begin{figure*}[!t]
\centering
\includegraphics[trim = 20mm 0mm 10mm 0mm, width=0.75\textwidth, height=0.45\textheight]{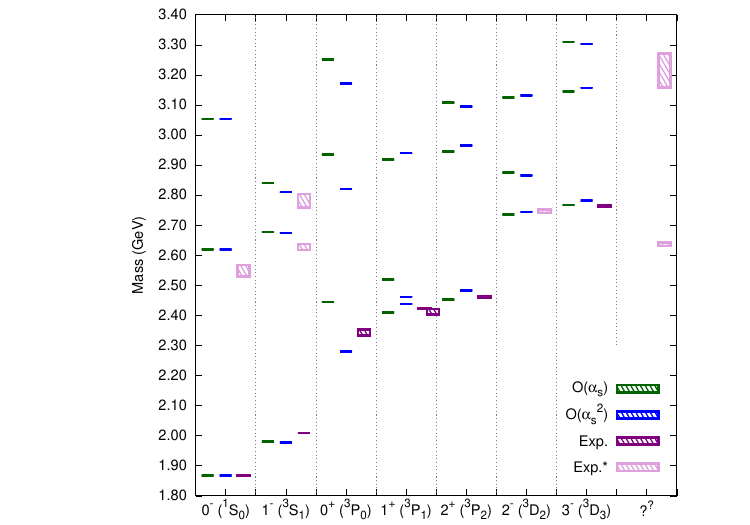}
\caption{\label{fig:spectrum} Charmed meson spectrum from constituent quark model (CQM) and experiment~\cite{ParticleDataGroup:2024cfk}. We show, for CQM's energy levels, the quark-antiquark value taking into account the one-gluon exchange potential ${\cal O}(\alpha_s)$ and including its one-loop correction ${\cal O}(\alpha_s^2)$. For experiment, we distinguish between well established states (purple bands) and those levels which still need confirmation and thus they have been omitted from the summary table (pink bands). The vertical extension of the experimental band is given by the experimental uncertainty.}
\end{figure*}

In summary, the \emph{na\"ive} constituent quark model is able to globally reproduce the spectrum of charmed mesons. In particular, there are higher-order terms of the gluon exchange potential that seem to be very significant in those channels of charmed mesons where there is a larger discrepancy between theory and experiment. As can be guessed from the discussion so far, and seen in Fig.~\ref{fig:spectrum}, when the next-to-leading (NLO) order term of the OGE interaction is included in the model, the spectrum of charmed mesons is described reasonably well. The numerical value of the scale-dependent effective strong coupling constant, $\alpha_s$, is $0.43$ in the $D$ meson sector. This implies that ${\cal O}(\alpha_s^2)$ one-gluon-exchange (OGE) potentials provide a mass correction of approximately $20\%$. The next-to-next-to-leading order (NNLO) corrections may introduce an additional $8\%$ variation in mass, which remains well within the theoretical uncertainties of the quark model and thus going beyond the scope of this manuscript.

It is worth noting herein that the discussion in the paragraph above does not mean that more complex structures such as tetraquark or meson-meson degrees of freedom in the meson's wave function cannot play a role but before resorting to them one should explore simpler refinements. Some exotic hadrons, such as $T_{cc}$, $T_{cs(\bar s)}$, and $Z_{c(b)}$, possess quantum numbers that unequivocally rule out a simple quark-antiquark interpretation, necessitating alternative descriptions such as multiquark configurations. These states do not couple to conventional mesons, implying that their dynamics must be understood beyond the na\"ive quark model, possibly as tetraquarks whose structure -- whether molecular or compact -- follows the same fundamental quark interactions~\cite{Ortega:2022efc, Ortega:2023azl, Ortega:2018cnm, Ortega:2021xst}. On the other hand, exotic hadrons like $X(3872)$ and $D_{s0}(2317)$ exhibit conventional quantum numbers but display characteristics that suggest a more complex nature. Their proximity to meson-meson thresholds leads to significant mass-shift renormalization, which must be accounted for when comparing their experimental masses to quark model predictions~\cite{Ortega:2009hj, Ortega:2016mms}. 

Finally, most of the original model parameters were constrained in the charmonium and bottomonium sectors and, as shown in Eq.~\eqref{eq:1loop}, there are one-loop OGE terms that impact in mesons made by an equally-flavor quark-antiquark pair. However, as one can see in Eq.~\eqref{eq:1loop}, the next-to-leading order corrections to the OGE potential are only tensor and spin-orbit terms, which are $1/(m_im_j)$ suppressed contributions, meaning that the $1$-loop corrections must be a factor $6$ smaller in the charmonium sector and a factor $50$ in the bottomonium one. Moreover, as one can see in Table~II and Fig.~1, the one-loop corrections to the original OGE potential produces mass shifts which are relatively small for most $D$-meson states. The largest effect is experienced by the $J^P=0^+$, $1^+$ and $2^+$ $D$ mesons, being notable for scalar states and just moderate for pseudo-vector and tensor states. Besides, in the charmonium sector, the operators $S_{ij}$ and $\vec{S}_{+}\cdot \vec{L}$ are active whereas the $\vec{S}_{-}\cdot \vec{L}$ operator is not, eliminating this additional effect. Therefore, the impact of the 1-loop OGE corrections in the charmonium sector is quite small, within the theoretical uncertainty of the CQM.


\section{SUMMARY}
\label{sec:summary}

We have evaluated the effectiveness of one-loop corrections to the one-gluon exchange potential in describing the spectrum of charmed mesons within a well-established constituent quark model. By incorporating these corrections, particularly spin-dependent terms that mainly affect $P$-wave states of mesons with different flavor quarks, the model aims to bridge gaps between theoretical predictions and experimental measurements across the charmed meson spectrum. The study investigates both well-established and recently observed states listed in the RPP of PDG.

The model successfully reproduces masses of many $S$- and $D$-wave states. Notably, $P$-wave states initially posed significant discrepancies with na\"ive quark model predictions. However, incorporating one-loop corrections to the OGE potential reduced these differences, aligning theoretical predictions more closely with observed values. This adjustment suggests that refinements of the na\"ive constituent quark model can be effectively reproduce the charmed meson spectrum without resorting first to exotic configurations, such as quark-gluon hybrids, compact tetraquarks or meson-meson molecules.

Overall, this enhanced CQM provides a refined framework for describing the heavy-light meson spectra, particularly offering insight into the nature of charmed mesons and the dynamics governing their mass structure. The results set a foundation for predicting higher-excited charmed states, potentially guiding future experimental searches and broadening the understanding of charmed meson interactions.




\begin{acknowledgments}
Work partially financed by 
the Escuela Polit\'ecnica Nacional under projects PIS-22-04 and PIM 23-01; 
EU Horizon 2020 research and innovation program, STRONG-2020 project, under grant agreement no. 824093;
Ministerio Espa\~nol de Ciencia e Innovaci\'on under grant nos. PID2022-141910NB-I00 and PID2022-140440NB-C22;
Junta de Castilla y Le\'on grant SA091P24 under program EDU/841/2024;
Junta de Andaluc\'ia under contract no. PAIDI FQM-370 and and PCI+D+i under the title: ``Tecnologías avanzadas para la exploración del universo y sus componentes” (Code AST22-0001).
\end{acknowledgments}


\bibliography{print_D-mesons}

\begin{thebibliography}{77}%
\makeatletter
\providecommand \@ifxundefined [1]{%
 \@ifx{#1\undefined}
}%
\providecommand \@ifnum [1]{%
 \ifnum #1\expandafter \@firstoftwo
 \else \expandafter \@secondoftwo
 \fi
}%
\providecommand \@ifx [1]{%
 \ifx #1\expandafter \@firstoftwo
 \else \expandafter \@secondoftwo
 \fi
}%
\providecommand \natexlab [1]{#1}%
\providecommand \enquote  [1]{``#1''}%
\providecommand \bibnamefont  [1]{#1}%
\providecommand \bibfnamefont [1]{#1}%
\providecommand \citenamefont [1]{#1}%
\providecommand \href@noop [0]{\@secondoftwo}%
\providecommand \href [0]{\begingroup \@sanitize@url \@href}%
\providecommand \@href[1]{\@@startlink{#1}\@@href}%
\providecommand \@@href[1]{\endgroup#1\@@endlink}%
\providecommand \@sanitize@url [0]{\catcode `\\12\catcode `\$12\catcode
  `\&12\catcode `\#12\catcode `\^12\catcode `\_12\catcode `\%12\relax}%
\providecommand \@@startlink[1]{}%
\providecommand \@@endlink[0]{}%
\providecommand \url  [0]{\begingroup\@sanitize@url \@url }%
\providecommand \@url [1]{\endgroup\@href {#1}{\urlprefix }}%
\providecommand \urlprefix  [0]{URL }%
\providecommand \Eprint [0]{\href }%
\providecommand \doibase [0]{https://doi.org/}%
\providecommand \selectlanguage [0]{\@gobble}%
\providecommand \bibinfo  [0]{\@secondoftwo}%
\providecommand \bibfield  [0]{\@secondoftwo}%
\providecommand \translation [1]{[#1]}%
\providecommand \BibitemOpen [0]{}%
\providecommand \bibitemStop [0]{}%
\providecommand \bibitemNoStop [0]{.\EOS\space}%
\providecommand \EOS [0]{\spacefactor3000\relax}%
\providecommand \BibitemShut  [1]{\csname bibitem#1\endcsname}%
\let\auto@bib@innerbib\@empty
\bibitem [{\citenamefont {Isgur}\ and\ \citenamefont
  {Wise}(1991)}]{Isgur:1991wq}%
  \BibitemOpen
  \bibfield  {author} {\bibinfo {author} {\bibfnamefont {N.}~\bibnamefont
  {Isgur}}\ and\ \bibinfo {author} {\bibfnamefont {M.~B.}\ \bibnamefont
  {Wise}},\ }\bibfield  {title} {\bibinfo {title} {{Spectroscopy with heavy
  quark symmetry}},\ }\href {https://doi.org/10.1103/PhysRevLett.66.1130}
  {\bibfield  {journal} {\bibinfo  {journal} {Phys. Rev. Lett.}\ }\textbf
  {\bibinfo {volume} {66}},\ \bibinfo {pages} {1130} (\bibinfo {year}
  {1991})}\BibitemShut {NoStop}%
\bibitem [{\citenamefont {Navas}\ \emph {et~al.}(2024)\citenamefont {Navas}
  \emph {et~al.}}]{ParticleDataGroup:2024cfk}%
  \BibitemOpen
  \bibfield  {author} {\bibinfo {author} {\bibfnamefont {S.}~\bibnamefont
  {Navas}} \emph {et~al.} (\bibinfo {collaboration} {Particle Data Group}),\
  }\bibfield  {title} {\bibinfo {title} {{Review of particle physics}},\ }\href
  {https://doi.org/10.1103/PhysRevD.110.030001} {\bibfield  {journal} {\bibinfo
   {journal} {Phys. Rev. D}\ }\textbf {\bibinfo {volume} {110}},\ \bibinfo
  {pages} {030001} (\bibinfo {year} {2024})}\BibitemShut {NoStop}%
\bibitem [{\citenamefont {Aaij}\ \emph
  {et~al.}(2015{\natexlab{a}})\citenamefont {Aaij} \emph
  {et~al.}}]{LHCb:2015klp}%
  \BibitemOpen
  \bibfield  {author} {\bibinfo {author} {\bibfnamefont {R.}~\bibnamefont
  {Aaij}} \emph {et~al.} (\bibinfo {collaboration} {LHCb}),\ }\bibfield
  {title} {\bibinfo {title} {{Dalitz plot analysis of $B^0 \to \overline{D}^0
  \pi^+\pi^-$ decays}},\ }\href {https://doi.org/10.1103/PhysRevD.92.032002}
  {\bibfield  {journal} {\bibinfo  {journal} {Phys. Rev. D}\ }\textbf {\bibinfo
  {volume} {92}},\ \bibinfo {pages} {032002} (\bibinfo {year}
  {2015}{\natexlab{a}})},\ \Eprint {https://arxiv.org/abs/1505.01710}
  {arXiv:1505.01710 [hep-ex]} \BibitemShut {NoStop}%
\bibitem [{\citenamefont {Aaij}\ \emph {et~al.}(2016)\citenamefont {Aaij} \emph
  {et~al.}}]{LHCb:2016lxy}%
  \BibitemOpen
  \bibfield  {author} {\bibinfo {author} {\bibfnamefont {R.}~\bibnamefont
  {Aaij}} \emph {et~al.} (\bibinfo {collaboration} {LHCb}),\ }\bibfield
  {title} {\bibinfo {title} {{Amplitude analysis of $B^{-} \to D^{+} \pi^{-}
  \pi^{-}$ decays}},\ }\href {https://doi.org/10.1103/PhysRevD.94.072001}
  {\bibfield  {journal} {\bibinfo  {journal} {Phys. Rev. D}\ }\textbf {\bibinfo
  {volume} {94}},\ \bibinfo {pages} {072001} (\bibinfo {year} {2016})},\
  \Eprint {https://arxiv.org/abs/1608.01289} {arXiv:1608.01289 [hep-ex]}
  \BibitemShut {NoStop}%
\bibitem [{\citenamefont {del Amo~Sanchez}\ \emph {et~al.}(2010)\citenamefont
  {del Amo~Sanchez} \emph {et~al.}}]{BaBar:2010zpy}%
  \BibitemOpen
  \bibfield  {author} {\bibinfo {author} {\bibfnamefont {P.}~\bibnamefont {del
  Amo~Sanchez}} \emph {et~al.} (\bibinfo {collaboration} {BaBar}),\ }\bibfield
  {title} {\bibinfo {title} {{Observation of new resonances decaying to $D\pi$
  and $D^*\pi$ in inclusive $e^+e^-$ collisions near $\sqrt{s}=$10.58 GeV}},\
  }\href {https://doi.org/10.1103/PhysRevD.82.111101} {\bibfield  {journal}
  {\bibinfo  {journal} {Phys. Rev. D}\ }\textbf {\bibinfo {volume} {82}},\
  \bibinfo {pages} {111101} (\bibinfo {year} {2010})},\ \Eprint
  {https://arxiv.org/abs/1009.2076} {arXiv:1009.2076 [hep-ex]} \BibitemShut
  {NoStop}%
\bibitem [{\citenamefont {Aaij}\ \emph {et~al.}(2013)\citenamefont {Aaij} \emph
  {et~al.}}]{LHCb:2013jjb}%
  \BibitemOpen
  \bibfield  {author} {\bibinfo {author} {\bibfnamefont {R.}~\bibnamefont
  {Aaij}} \emph {et~al.} (\bibinfo {collaboration} {LHCb}),\ }\bibfield
  {title} {\bibinfo {title} {{Study of $D_J$ meson decays to $D^+\pi^-$, $D^0
  \pi^+$ and $D^{*+}\pi^-$ final states in pp collision}},\ }\href
  {https://doi.org/10.1007/JHEP09(2013)145} {\bibfield  {journal} {\bibinfo
  {journal} {JHEP}\ }\textbf {\bibinfo {volume} {09}},\ \bibinfo {pages}
  {145}},\ \Eprint {https://arxiv.org/abs/1307.4556} {arXiv:1307.4556 [hep-ex]}
  \BibitemShut {NoStop}%
\bibitem [{\citenamefont {Aaij}\ \emph
  {et~al.}(2015{\natexlab{b}})\citenamefont {Aaij} \emph
  {et~al.}}]{LHCb:2015eqv}%
  \BibitemOpen
  \bibfield  {author} {\bibinfo {author} {\bibfnamefont {R.}~\bibnamefont
  {Aaij}} \emph {et~al.} (\bibinfo {collaboration} {LHCb}),\ }\bibfield
  {title} {\bibinfo {title} {{First observation and amplitude analysis of the
  $B^{-}\to D^{+}K^{-}\pi^{-}$ decay}},\ }\href
  {https://doi.org/10.1103/PhysRevD.91.092002} {\bibfield  {journal} {\bibinfo
  {journal} {Phys. Rev. D}\ }\textbf {\bibinfo {volume} {91}},\ \bibinfo
  {pages} {092002} (\bibinfo {year} {2015}{\natexlab{b}})},\ \bibinfo {note}
  {[Erratum: Phys.Rev.D 93, 119901 (2016)]},\ \Eprint
  {https://arxiv.org/abs/1503.02995} {arXiv:1503.02995 [hep-ex]} \BibitemShut
  {NoStop}%
\bibitem [{\citenamefont {Abreu}\ \emph {et~al.}(1998)\citenamefont {Abreu}
  \emph {et~al.}}]{DELPHI:1998oyl}%
  \BibitemOpen
  \bibfield  {author} {\bibinfo {author} {\bibfnamefont {P.}~\bibnamefont
  {Abreu}} \emph {et~al.} (\bibinfo {collaboration} {DELPHI}),\ }\bibfield
  {title} {\bibinfo {title} {{First evidence for a charm radial excitation,
  D*-prime}},\ }\href {https://doi.org/10.1016/S0370-2693(98)00346-3}
  {\bibfield  {journal} {\bibinfo  {journal} {Phys. Lett. B}\ }\textbf
  {\bibinfo {volume} {426}},\ \bibinfo {pages} {231} (\bibinfo {year}
  {1998})}\BibitemShut {NoStop}%
\bibitem [{\citenamefont {Abbiendi}\ \emph {et~al.}(2001)\citenamefont
  {Abbiendi} \emph {et~al.}}]{OPAL:2001xfs}%
  \BibitemOpen
  \bibfield  {author} {\bibinfo {author} {\bibfnamefont {G.}~\bibnamefont
  {Abbiendi}} \emph {et~al.} (\bibinfo {collaboration} {OPAL}),\ }\bibfield
  {title} {\bibinfo {title} {{A Search for a narrow radial excitation of the
  D*+- meson}},\ }\href {https://doi.org/10.1007/s100520100696} {\bibfield
  {journal} {\bibinfo  {journal} {Eur. Phys. J. C}\ }\textbf {\bibinfo {volume}
  {20}},\ \bibinfo {pages} {445} (\bibinfo {year} {2001})},\ \Eprint
  {https://arxiv.org/abs/hep-ex/0101045} {arXiv:hep-ex/0101045} \BibitemShut
  {NoStop}%
\bibitem [{\citenamefont {Chekanov}\ \emph {et~al.}(2009)\citenamefont
  {Chekanov} \emph {et~al.}}]{ZEUS:2008nzg}%
  \BibitemOpen
  \bibfield  {author} {\bibinfo {author} {\bibfnamefont {S.}~\bibnamefont
  {Chekanov}} \emph {et~al.} (\bibinfo {collaboration} {ZEUS}),\ }\bibfield
  {title} {\bibinfo {title} {{Production of excited charm and charm-strange
  mesons at HERA}},\ }\href {https://doi.org/10.1140/epjc/s10052-009-0881-x}
  {\bibfield  {journal} {\bibinfo  {journal} {Eur. Phys. J. C}\ }\textbf
  {\bibinfo {volume} {60}},\ \bibinfo {pages} {25} (\bibinfo {year} {2009})},\
  \Eprint {https://arxiv.org/abs/0807.1290} {arXiv:0807.1290 [hep-ex]}
  \BibitemShut {NoStop}%
\bibitem [{\citenamefont {Godfrey}\ and\ \citenamefont
  {Isgur}(1985)}]{Godfrey:1985xj}%
  \BibitemOpen
  \bibfield  {author} {\bibinfo {author} {\bibfnamefont {S.}~\bibnamefont
  {Godfrey}}\ and\ \bibinfo {author} {\bibfnamefont {N.}~\bibnamefont
  {Isgur}},\ }\bibfield  {title} {\bibinfo {title} {{Mesons in a Relativized
  Quark Model with Chromodynamics}},\ }\href
  {https://doi.org/10.1103/PhysRevD.32.189} {\bibfield  {journal} {\bibinfo
  {journal} {Phys. Rev. D}\ }\textbf {\bibinfo {volume} {32}},\ \bibinfo
  {pages} {189} (\bibinfo {year} {1985})}\BibitemShut {NoStop}%
\bibitem [{\citenamefont {Zeng}\ \emph {et~al.}(1995)\citenamefont {Zeng},
  \citenamefont {Van~Orden},\ and\ \citenamefont {Roberts}}]{Zeng:1994vj}%
  \BibitemOpen
  \bibfield  {author} {\bibinfo {author} {\bibfnamefont {J.}~\bibnamefont
  {Zeng}}, \bibinfo {author} {\bibfnamefont {J.~W.}\ \bibnamefont
  {Van~Orden}},\ and\ \bibinfo {author} {\bibfnamefont {W.}~\bibnamefont
  {Roberts}},\ }\bibfield  {title} {\bibinfo {title} {{Heavy mesons in a
  relativistic model}},\ }\href {https://doi.org/10.1103/PhysRevD.52.5229}
  {\bibfield  {journal} {\bibinfo  {journal} {Phys. Rev. D}\ }\textbf {\bibinfo
  {volume} {52}},\ \bibinfo {pages} {5229} (\bibinfo {year} {1995})},\ \Eprint
  {https://arxiv.org/abs/hep-ph/9412269} {arXiv:hep-ph/9412269} \BibitemShut
  {NoStop}%
\bibitem [{\citenamefont {Gupta}\ and\ \citenamefont
  {Johnson}(1995)}]{Gupta:1994mw}%
  \BibitemOpen
  \bibfield  {author} {\bibinfo {author} {\bibfnamefont {S.~N.}\ \bibnamefont
  {Gupta}}\ and\ \bibinfo {author} {\bibfnamefont {J.~M.}\ \bibnamefont
  {Johnson}},\ }\bibfield  {title} {\bibinfo {title} {{Quantum chromodynamic
  potential model for light heavy quarkonia and the heavy quark effective
  theory}},\ }\href {https://doi.org/10.1103/PhysRevD.51.168} {\bibfield
  {journal} {\bibinfo  {journal} {Phys. Rev. D}\ }\textbf {\bibinfo {volume}
  {51}},\ \bibinfo {pages} {168} (\bibinfo {year} {1995})},\ \Eprint
  {https://arxiv.org/abs/hep-ph/9409432} {arXiv:hep-ph/9409432} \BibitemShut
  {NoStop}%
\bibitem [{\citenamefont {Lahde}\ \emph {et~al.}(2000)\citenamefont {Lahde},
  \citenamefont {Nyfalt},\ and\ \citenamefont {Riska}}]{Lahde:1999ih}%
  \BibitemOpen
  \bibfield  {author} {\bibinfo {author} {\bibfnamefont {T.~A.}\ \bibnamefont
  {Lahde}}, \bibinfo {author} {\bibfnamefont {C.~J.}\ \bibnamefont {Nyfalt}},\
  and\ \bibinfo {author} {\bibfnamefont {D.~O.}\ \bibnamefont {Riska}},\
  }\bibfield  {title} {\bibinfo {title} {{Spectra and M1 decay widths of heavy
  light mesons}},\ }\href {https://doi.org/10.1016/S0375-9474(00)00154-8}
  {\bibfield  {journal} {\bibinfo  {journal} {Nucl. Phys. A}\ }\textbf
  {\bibinfo {volume} {674}},\ \bibinfo {pages} {141} (\bibinfo {year}
  {2000})},\ \Eprint {https://arxiv.org/abs/hep-ph/9908485}
  {arXiv:hep-ph/9908485} \BibitemShut {NoStop}%
\bibitem [{\citenamefont {Mohler}\ and\ \citenamefont
  {Woloshyn}(2011)}]{Mohler:2011ke}%
  \BibitemOpen
  \bibfield  {author} {\bibinfo {author} {\bibfnamefont {D.}~\bibnamefont
  {Mohler}}\ and\ \bibinfo {author} {\bibfnamefont {R.~M.}\ \bibnamefont
  {Woloshyn}},\ }\bibfield  {title} {\bibinfo {title} {{$D$ and $D_s$ meson
  spectroscopy}},\ }\href {https://doi.org/10.1103/PhysRevD.84.054505}
  {\bibfield  {journal} {\bibinfo  {journal} {Phys. Rev. D}\ }\textbf {\bibinfo
  {volume} {84}},\ \bibinfo {pages} {054505} (\bibinfo {year} {2011})},\
  \Eprint {https://arxiv.org/abs/1103.5506} {arXiv:1103.5506 [hep-lat]}
  \BibitemShut {NoStop}%
\bibitem [{\citenamefont {Moir}\ \emph {et~al.}(2013)\citenamefont {Moir},
  \citenamefont {Peardon}, \citenamefont {Ryan}, \citenamefont {Thomas},\ and\
  \citenamefont {Liu}}]{Moir:2013ub}%
  \BibitemOpen
  \bibfield  {author} {\bibinfo {author} {\bibfnamefont {G.}~\bibnamefont
  {Moir}}, \bibinfo {author} {\bibfnamefont {M.}~\bibnamefont {Peardon}},
  \bibinfo {author} {\bibfnamefont {S.~M.}\ \bibnamefont {Ryan}}, \bibinfo
  {author} {\bibfnamefont {C.~E.}\ \bibnamefont {Thomas}},\ and\ \bibinfo
  {author} {\bibfnamefont {L.}~\bibnamefont {Liu}},\ }\bibfield  {title}
  {\bibinfo {title} {{Excited spectroscopy of charmed mesons from lattice
  QCD}},\ }\href {https://doi.org/10.1007/JHEP05(2013)021} {\bibfield
  {journal} {\bibinfo  {journal} {JHEP}\ }\textbf {\bibinfo {volume} {05}},\
  \bibinfo {pages} {021}},\ \Eprint {https://arxiv.org/abs/1301.7670}
  {arXiv:1301.7670 [hep-ph]} \BibitemShut {NoStop}%
\bibitem [{\citenamefont {Kalinowski}\ and\ \citenamefont
  {Wagner}(2015)}]{Kalinowski:2015bwa}%
  \BibitemOpen
  \bibfield  {author} {\bibinfo {author} {\bibfnamefont {M.}~\bibnamefont
  {Kalinowski}}\ and\ \bibinfo {author} {\bibfnamefont {M.}~\bibnamefont
  {Wagner}},\ }\bibfield  {title} {\bibinfo {title} {{Masses of $D$ mesons,
  $D_s$ mesons and charmonium states from twisted mass lattice QCD}},\ }\href
  {https://doi.org/10.1103/PhysRevD.92.094508} {\bibfield  {journal} {\bibinfo
  {journal} {Phys. Rev. D}\ }\textbf {\bibinfo {volume} {92}},\ \bibinfo
  {pages} {094508} (\bibinfo {year} {2015})},\ \Eprint
  {https://arxiv.org/abs/1509.02396} {arXiv:1509.02396 [hep-lat]} \BibitemShut
  {NoStop}%
\bibitem [{\citenamefont {Cichy}\ \emph {et~al.}(2016)\citenamefont {Cichy},
  \citenamefont {Kalinowski},\ and\ \citenamefont {Wagner}}]{Cichy:2016bci}%
  \BibitemOpen
  \bibfield  {author} {\bibinfo {author} {\bibfnamefont {K.}~\bibnamefont
  {Cichy}}, \bibinfo {author} {\bibfnamefont {M.}~\bibnamefont {Kalinowski}},\
  and\ \bibinfo {author} {\bibfnamefont {M.}~\bibnamefont {Wagner}},\
  }\bibfield  {title} {\bibinfo {title} {{Continuum limit of the $D$ meson,
  $D_s$ meson and charmonium spectrum from $N_f=2+1+1$ twisted mass lattice
  QCD}},\ }\href {https://doi.org/10.1103/PhysRevD.94.094503} {\bibfield
  {journal} {\bibinfo  {journal} {Phys. Rev. D}\ }\textbf {\bibinfo {volume}
  {94}},\ \bibinfo {pages} {094503} (\bibinfo {year} {2016})},\ \Eprint
  {https://arxiv.org/abs/1603.06467} {arXiv:1603.06467 [hep-lat]} \BibitemShut
  {NoStop}%
\bibitem [{\citenamefont {Gayer}\ \emph {et~al.}(2021)\citenamefont {Gayer},
  \citenamefont {Lang}, \citenamefont {Ryan}, \citenamefont {Tims},
  \citenamefont {Thomas},\ and\ \citenamefont {Wilson}}]{Gayer:2021xzv}%
  \BibitemOpen
  \bibfield  {author} {\bibinfo {author} {\bibfnamefont {L.}~\bibnamefont
  {Gayer}}, \bibinfo {author} {\bibfnamefont {N.}~\bibnamefont {Lang}},
  \bibinfo {author} {\bibfnamefont {S.~M.}\ \bibnamefont {Ryan}}, \bibinfo
  {author} {\bibfnamefont {D.}~\bibnamefont {Tims}}, \bibinfo {author}
  {\bibfnamefont {C.~E.}\ \bibnamefont {Thomas}},\ and\ \bibinfo {author}
  {\bibfnamefont {D.~J.}\ \bibnamefont {Wilson}} (\bibinfo {collaboration}
  {Hadron Spectrum}),\ }\bibfield  {title} {\bibinfo {title} {{Isospin-1/2
  D\ensuremath{\pi} scattering and the lightest $ {D}_0^{\ast } $ resonance
  from lattice QCD}},\ }\href {https://doi.org/10.1007/JHEP07(2021)123}
  {\bibfield  {journal} {\bibinfo  {journal} {JHEP}\ }\textbf {\bibinfo
  {volume} {07}},\ \bibinfo {pages} {123}},\ \Eprint
  {https://arxiv.org/abs/2102.04973} {arXiv:2102.04973 [hep-lat]} \BibitemShut
  {NoStop}%
\bibitem [{\citenamefont {Albaladejo}\ \emph {et~al.}(2017)\citenamefont
  {Albaladejo}, \citenamefont {Fernandez-Soler}, \citenamefont {Guo},\ and\
  \citenamefont {Nieves}}]{Albaladejo:2016lbb}%
  \BibitemOpen
  \bibfield  {author} {\bibinfo {author} {\bibfnamefont {M.}~\bibnamefont
  {Albaladejo}}, \bibinfo {author} {\bibfnamefont {P.}~\bibnamefont
  {Fernandez-Soler}}, \bibinfo {author} {\bibfnamefont {F.-K.}\ \bibnamefont
  {Guo}},\ and\ \bibinfo {author} {\bibfnamefont {J.}~\bibnamefont {Nieves}},\
  }\bibfield  {title} {\bibinfo {title} {{Two-pole structure of the
  $D^\ast_0(2400)$}},\ }\href {https://doi.org/10.1016/j.physletb.2017.02.036}
  {\bibfield  {journal} {\bibinfo  {journal} {Phys. Lett. B}\ }\textbf
  {\bibinfo {volume} {767}},\ \bibinfo {pages} {465} (\bibinfo {year}
  {2017})},\ \Eprint {https://arxiv.org/abs/1610.06727} {arXiv:1610.06727
  [hep-ph]} \BibitemShut {NoStop}%
\bibitem [{\citenamefont {Du}\ \emph {et~al.}(2018)\citenamefont {Du},
  \citenamefont {Albaladejo}, \citenamefont {Fern\'andez-Soler}, \citenamefont
  {Guo}, \citenamefont {Hanhart}, \citenamefont {Mei\ss{}ner}, \citenamefont
  {Nieves},\ and\ \citenamefont {Yao}}]{Du:2017zvv}%
  \BibitemOpen
  \bibfield  {author} {\bibinfo {author} {\bibfnamefont {M.-L.}\ \bibnamefont
  {Du}}, \bibinfo {author} {\bibfnamefont {M.}~\bibnamefont {Albaladejo}},
  \bibinfo {author} {\bibfnamefont {P.}~\bibnamefont {Fern\'andez-Soler}},
  \bibinfo {author} {\bibfnamefont {F.-K.}\ \bibnamefont {Guo}}, \bibinfo
  {author} {\bibfnamefont {C.}~\bibnamefont {Hanhart}}, \bibinfo {author}
  {\bibfnamefont {U.-G.}\ \bibnamefont {Mei\ss{}ner}}, \bibinfo {author}
  {\bibfnamefont {J.}~\bibnamefont {Nieves}},\ and\ \bibinfo {author}
  {\bibfnamefont {D.-L.}\ \bibnamefont {Yao}},\ }\bibfield  {title} {\bibinfo
  {title} {{Towards a new paradigm for heavy-light meson spectroscopy}},\
  }\href {https://doi.org/10.1103/PhysRevD.98.094018} {\bibfield  {journal}
  {\bibinfo  {journal} {Phys. Rev. D}\ }\textbf {\bibinfo {volume} {98}},\
  \bibinfo {pages} {094018} (\bibinfo {year} {2018})},\ \Eprint
  {https://arxiv.org/abs/1712.07957} {arXiv:1712.07957 [hep-ph]} \BibitemShut
  {NoStop}%
\bibitem [{\citenamefont {Du}\ \emph {et~al.}(2021)\citenamefont {Du},
  \citenamefont {Guo}, \citenamefont {Hanhart}, \citenamefont {Kubis},\ and\
  \citenamefont {Mei\ss{}ner}}]{Du:2020pui}%
  \BibitemOpen
  \bibfield  {author} {\bibinfo {author} {\bibfnamefont {M.-L.}\ \bibnamefont
  {Du}}, \bibinfo {author} {\bibfnamefont {F.-K.}\ \bibnamefont {Guo}},
  \bibinfo {author} {\bibfnamefont {C.}~\bibnamefont {Hanhart}}, \bibinfo
  {author} {\bibfnamefont {B.}~\bibnamefont {Kubis}},\ and\ \bibinfo {author}
  {\bibfnamefont {U.-G.}\ \bibnamefont {Mei\ss{}ner}},\ }\bibfield  {title}
  {\bibinfo {title} {{Where is the lightest charmed scalar meson?}},\ }\href
  {https://doi.org/10.1103/PhysRevLett.126.192001} {\bibfield  {journal}
  {\bibinfo  {journal} {Phys. Rev. Lett.}\ }\textbf {\bibinfo {volume} {126}},\
  \bibinfo {pages} {192001} (\bibinfo {year} {2021})},\ \Eprint
  {https://arxiv.org/abs/2012.04599} {arXiv:2012.04599 [hep-ph]} \BibitemShut
  {NoStop}%
\bibitem [{\citenamefont {Wang}(2011)}]{Wang:2010ydc}%
  \BibitemOpen
  \bibfield  {author} {\bibinfo {author} {\bibfnamefont {Z.-G.}\ \bibnamefont
  {Wang}},\ }\bibfield  {title} {\bibinfo {title} {{Analysis of strong decays
  of the charmed mesons D(2550), D(2600), D(2750) and D(2760)}},\ }\href
  {https://doi.org/10.1103/PhysRevD.83.014009} {\bibfield  {journal} {\bibinfo
  {journal} {Phys. Rev. D}\ }\textbf {\bibinfo {volume} {83}},\ \bibinfo
  {pages} {014009} (\bibinfo {year} {2011})},\ \Eprint
  {https://arxiv.org/abs/1009.3605} {arXiv:1009.3605 [hep-ph]} \BibitemShut
  {NoStop}%
\bibitem [{\citenamefont {Wang}(2013)}]{Wang:2013tka}%
  \BibitemOpen
  \bibfield  {author} {\bibinfo {author} {\bibfnamefont {Z.-G.}\ \bibnamefont
  {Wang}},\ }\bibfield  {title} {\bibinfo {title} {{Analysis of strong decays
  of the charmed mesons $D_J(2580), D^*_J(2650), D_J(2740), D^*_J(2760),
  D_J(3000), D^*_J(3000)$}},\ }\href
  {https://doi.org/10.1103/PhysRevD.88.114003} {\bibfield  {journal} {\bibinfo
  {journal} {Phys. Rev. D}\ }\textbf {\bibinfo {volume} {88}},\ \bibinfo
  {pages} {114003} (\bibinfo {year} {2013})},\ \Eprint
  {https://arxiv.org/abs/1308.0533} {arXiv:1308.0533 [hep-ph]} \BibitemShut
  {NoStop}%
\bibitem [{\citenamefont {Li}\ \emph {et~al.}(2007)\citenamefont {Li},
  \citenamefont {Ma},\ and\ \citenamefont {Liu}}]{Li:2007px}%
  \BibitemOpen
  \bibfield  {author} {\bibinfo {author} {\bibfnamefont {D.-M.}\ \bibnamefont
  {Li}}, \bibinfo {author} {\bibfnamefont {B.}~\bibnamefont {Ma}},\ and\
  \bibinfo {author} {\bibfnamefont {Y.-H.}\ \bibnamefont {Liu}},\ }\bibfield
  {title} {\bibinfo {title} {{Understanding masses of c anti-s states in Regge
  phenomenology}},\ }\href {https://doi.org/10.1140/epjc/s10052-007-0286-7}
  {\bibfield  {journal} {\bibinfo  {journal} {Eur. Phys. J. C}\ }\textbf
  {\bibinfo {volume} {51}},\ \bibinfo {pages} {359} (\bibinfo {year} {2007})},\
  \Eprint {https://arxiv.org/abs/hep-ph/0703278} {arXiv:hep-ph/0703278}
  \BibitemShut {NoStop}%
\bibitem [{\citenamefont {Chen}(2018)}]{Chen:2018nnr}%
  \BibitemOpen
  \bibfield  {author} {\bibinfo {author} {\bibfnamefont {J.-K.}\ \bibnamefont
  {Chen}},\ }\bibfield  {title} {\bibinfo {title} {{Regge trajectories for the
  mesons consisting of different quarks}},\ }\href
  {https://doi.org/10.1140/epjc/s10052-018-6134-0} {\bibfield  {journal}
  {\bibinfo  {journal} {Eur. Phys. J. C}\ }\textbf {\bibinfo {volume} {78}},\
  \bibinfo {pages} {648} (\bibinfo {year} {2018})}\BibitemShut {NoStop}%
\bibitem [{\citenamefont {Di~Pierro}\ and\ \citenamefont
  {Eichten}(2001)}]{DiPierro:2001dwf}%
  \BibitemOpen
  \bibfield  {author} {\bibinfo {author} {\bibfnamefont {M.}~\bibnamefont
  {Di~Pierro}}\ and\ \bibinfo {author} {\bibfnamefont {E.}~\bibnamefont
  {Eichten}},\ }\bibfield  {title} {\bibinfo {title} {{Excited Heavy - Light
  Systems and Hadronic Transitions}},\ }\href
  {https://doi.org/10.1103/PhysRevD.64.114004} {\bibfield  {journal} {\bibinfo
  {journal} {Phys. Rev. D}\ }\textbf {\bibinfo {volume} {64}},\ \bibinfo
  {pages} {114004} (\bibinfo {year} {2001})},\ \Eprint
  {https://arxiv.org/abs/hep-ph/0104208} {arXiv:hep-ph/0104208} \BibitemShut
  {NoStop}%
\bibitem [{\citenamefont {Godfrey}(2005)}]{Godfrey:2005ww}%
  \BibitemOpen
  \bibfield  {author} {\bibinfo {author} {\bibfnamefont {S.}~\bibnamefont
  {Godfrey}},\ }\bibfield  {title} {\bibinfo {title} {{Properties of the
  charmed P-wave mesons}},\ }\href {https://doi.org/10.1103/PhysRevD.72.054029}
  {\bibfield  {journal} {\bibinfo  {journal} {Phys. Rev. D}\ }\textbf {\bibinfo
  {volume} {72}},\ \bibinfo {pages} {054029} (\bibinfo {year} {2005})},\
  \Eprint {https://arxiv.org/abs/hep-ph/0508078} {arXiv:hep-ph/0508078}
  \BibitemShut {NoStop}%
\bibitem [{\citenamefont {Close}\ and\ \citenamefont
  {Swanson}(2005)}]{Close:2005se}%
  \BibitemOpen
  \bibfield  {author} {\bibinfo {author} {\bibfnamefont {F.~E.}\ \bibnamefont
  {Close}}\ and\ \bibinfo {author} {\bibfnamefont {E.~S.}\ \bibnamefont
  {Swanson}},\ }\bibfield  {title} {\bibinfo {title} {{Dynamics and decay of
  heavy-light hadrons}},\ }\href {https://doi.org/10.1103/PhysRevD.72.094004}
  {\bibfield  {journal} {\bibinfo  {journal} {Phys. Rev. D}\ }\textbf {\bibinfo
  {volume} {72}},\ \bibinfo {pages} {094004} (\bibinfo {year} {2005})},\
  \Eprint {https://arxiv.org/abs/hep-ph/0505206} {arXiv:hep-ph/0505206}
  \BibitemShut {NoStop}%
\bibitem [{\citenamefont {Close}\ \emph {et~al.}(2007)\citenamefont {Close},
  \citenamefont {Thomas}, \citenamefont {Lakhina},\ and\ \citenamefont
  {Swanson}}]{Close:2006gr}%
  \BibitemOpen
  \bibfield  {author} {\bibinfo {author} {\bibfnamefont {F.~E.}\ \bibnamefont
  {Close}}, \bibinfo {author} {\bibfnamefont {C.~E.}\ \bibnamefont {Thomas}},
  \bibinfo {author} {\bibfnamefont {O.}~\bibnamefont {Lakhina}},\ and\ \bibinfo
  {author} {\bibfnamefont {E.~S.}\ \bibnamefont {Swanson}},\ }\bibfield
  {title} {\bibinfo {title} {{Canonical interpretation of the D(sJ)(2860) and
  D(sJ)(2690)}},\ }\href {https://doi.org/10.1016/j.physletb.2007.01.052}
  {\bibfield  {journal} {\bibinfo  {journal} {Phys. Lett. B}\ }\textbf
  {\bibinfo {volume} {647}},\ \bibinfo {pages} {159} (\bibinfo {year}
  {2007})},\ \Eprint {https://arxiv.org/abs/hep-ph/0608139}
  {arXiv:hep-ph/0608139} \BibitemShut {NoStop}%
\bibitem [{\citenamefont {Zhang}\ \emph {et~al.}(2007)\citenamefont {Zhang},
  \citenamefont {Liu}, \citenamefont {Deng},\ and\ \citenamefont
  {Zhu}}]{Zhang:2006yj}%
  \BibitemOpen
  \bibfield  {author} {\bibinfo {author} {\bibfnamefont {B.}~\bibnamefont
  {Zhang}}, \bibinfo {author} {\bibfnamefont {X.}~\bibnamefont {Liu}}, \bibinfo
  {author} {\bibfnamefont {W.-Z.}\ \bibnamefont {Deng}},\ and\ \bibinfo
  {author} {\bibfnamefont {S.-L.}\ \bibnamefont {Zhu}},\ }\bibfield  {title}
  {\bibinfo {title} {{$D_ {sJ} (2860)$ and $D_ {sJ} (2715)$}},\ }\href
  {https://doi.org/10.1140/epjc/s10052-007-0221-y} {\bibfield  {journal}
  {\bibinfo  {journal} {Eur. Phys. J. C}\ }\textbf {\bibinfo {volume} {50}},\
  \bibinfo {pages} {617} (\bibinfo {year} {2007})},\ \Eprint
  {https://arxiv.org/abs/hep-ph/0609013} {arXiv:hep-ph/0609013} \BibitemShut
  {NoStop}%
\bibitem [{\citenamefont {Wei}\ \emph {et~al.}(2007)\citenamefont {Wei},
  \citenamefont {Liu},\ and\ \citenamefont {Zhu}}]{Wei:2006wa}%
  \BibitemOpen
  \bibfield  {author} {\bibinfo {author} {\bibfnamefont {W.}~\bibnamefont
  {Wei}}, \bibinfo {author} {\bibfnamefont {X.}~\bibnamefont {Liu}},\ and\
  \bibinfo {author} {\bibfnamefont {S.-L.}\ \bibnamefont {Zhu}},\ }\bibfield
  {title} {\bibinfo {title} {{D wave heavy mesons}},\ }\href
  {https://doi.org/10.1103/PhysRevD.75.014013} {\bibfield  {journal} {\bibinfo
  {journal} {Phys. Rev. D}\ }\textbf {\bibinfo {volume} {75}},\ \bibinfo
  {pages} {014013} (\bibinfo {year} {2007})},\ \Eprint
  {https://arxiv.org/abs/hep-ph/0612066} {arXiv:hep-ph/0612066} \BibitemShut
  {NoStop}%
\bibitem [{\citenamefont {Ebert}\ \emph {et~al.}(2010)\citenamefont {Ebert},
  \citenamefont {Faustov},\ and\ \citenamefont {Galkin}}]{Ebert:2009ua}%
  \BibitemOpen
  \bibfield  {author} {\bibinfo {author} {\bibfnamefont {D.}~\bibnamefont
  {Ebert}}, \bibinfo {author} {\bibfnamefont {R.~N.}\ \bibnamefont {Faustov}},\
  and\ \bibinfo {author} {\bibfnamefont {V.~O.}\ \bibnamefont {Galkin}},\
  }\bibfield  {title} {\bibinfo {title} {{Heavy-light meson spectroscopy and
  Regge trajectories in the relativistic quark model}},\ }\href
  {https://doi.org/10.1140/epjc/s10052-010-1233-6} {\bibfield  {journal}
  {\bibinfo  {journal} {Eur. Phys. J. C}\ }\textbf {\bibinfo {volume} {66}},\
  \bibinfo {pages} {197} (\bibinfo {year} {2010})},\ \Eprint
  {https://arxiv.org/abs/0910.5612} {arXiv:0910.5612 [hep-ph]} \BibitemShut
  {NoStop}%
\bibitem [{\citenamefont {Song}\ \emph {et~al.}(2015)\citenamefont {Song},
  \citenamefont {Chen}, \citenamefont {Liu},\ and\ \citenamefont
  {Matsuki}}]{Song:2015fha}%
  \BibitemOpen
  \bibfield  {author} {\bibinfo {author} {\bibfnamefont {Q.-T.}\ \bibnamefont
  {Song}}, \bibinfo {author} {\bibfnamefont {D.-Y.}\ \bibnamefont {Chen}},
  \bibinfo {author} {\bibfnamefont {X.}~\bibnamefont {Liu}},\ and\ \bibinfo
  {author} {\bibfnamefont {T.}~\bibnamefont {Matsuki}},\ }\bibfield  {title}
  {\bibinfo {title} {{Higher radial and orbital excitations in the charmed
  meson family}},\ }\href {https://doi.org/10.1103/PhysRevD.92.074011}
  {\bibfield  {journal} {\bibinfo  {journal} {Phys. Rev. D}\ }\textbf {\bibinfo
  {volume} {92}},\ \bibinfo {pages} {074011} (\bibinfo {year} {2015})},\
  \Eprint {https://arxiv.org/abs/1503.05728} {arXiv:1503.05728 [hep-ph]}
  \BibitemShut {NoStop}%
\bibitem [{\citenamefont {Ferretti}\ and\ \citenamefont
  {Santopinto}(2018)}]{Ferretti:2015rsa}%
  \BibitemOpen
  \bibfield  {author} {\bibinfo {author} {\bibfnamefont {J.}~\bibnamefont
  {Ferretti}}\ and\ \bibinfo {author} {\bibfnamefont {E.}~\bibnamefont
  {Santopinto}},\ }\bibfield  {title} {\bibinfo {title} {{Open-flavor strong
  decays of open-charm and open-bottom mesons in the $^3P_0$ model}},\ }\href
  {https://doi.org/10.1103/PhysRevD.97.114020} {\bibfield  {journal} {\bibinfo
  {journal} {Phys. Rev. D}\ }\textbf {\bibinfo {volume} {97}},\ \bibinfo
  {pages} {114020} (\bibinfo {year} {2018})},\ \Eprint
  {https://arxiv.org/abs/1506.04415} {arXiv:1506.04415 [hep-ph]} \BibitemShut
  {NoStop}%
\bibitem [{\citenamefont {Liu}\ and\ \citenamefont {Lu}(2017)}]{Liu:2016efm}%
  \BibitemOpen
  \bibfield  {author} {\bibinfo {author} {\bibfnamefont {J.-B.}\ \bibnamefont
  {Liu}}\ and\ \bibinfo {author} {\bibfnamefont {C.-D.}\ \bibnamefont {Lu}},\
  }\bibfield  {title} {\bibinfo {title} {{Spectra of heavy\textendash{}light
  mesons in a relativistic model}},\ }\href
  {https://doi.org/10.1140/epjc/s10052-017-4867-9} {\bibfield  {journal}
  {\bibinfo  {journal} {Eur. Phys. J. C}\ }\textbf {\bibinfo {volume} {77}},\
  \bibinfo {pages} {312} (\bibinfo {year} {2017})},\ \Eprint
  {https://arxiv.org/abs/1605.05550} {arXiv:1605.05550 [hep-ph]} \BibitemShut
  {NoStop}%
\bibitem [{\citenamefont {Barnes}\ \emph {et~al.}(2003)\citenamefont {Barnes},
  \citenamefont {Close},\ and\ \citenamefont {Lipkin}}]{Barnes:2003dj}%
  \BibitemOpen
  \bibfield  {author} {\bibinfo {author} {\bibfnamefont {T.}~\bibnamefont
  {Barnes}}, \bibinfo {author} {\bibfnamefont {F.~E.}\ \bibnamefont {Close}},\
  and\ \bibinfo {author} {\bibfnamefont {H.~J.}\ \bibnamefont {Lipkin}},\
  }\bibfield  {title} {\bibinfo {title} {{Implications of a DK molecule at
  2.32-GeV}},\ }\href {https://doi.org/10.1103/PhysRevD.68.054006} {\bibfield
  {journal} {\bibinfo  {journal} {Phys. Rev. D}\ }\textbf {\bibinfo {volume}
  {68}},\ \bibinfo {pages} {054006} (\bibinfo {year} {2003})},\ \Eprint
  {https://arxiv.org/abs/hep-ph/0305025} {arXiv:hep-ph/0305025} \BibitemShut
  {NoStop}%
\bibitem [{\citenamefont {van Beveren}\ and\ \citenamefont
  {Rupp}(2003)}]{vanBeveren:2003kd}%
  \BibitemOpen
  \bibfield  {author} {\bibinfo {author} {\bibfnamefont {E.}~\bibnamefont {van
  Beveren}}\ and\ \bibinfo {author} {\bibfnamefont {G.}~\bibnamefont {Rupp}},\
  }\bibfield  {title} {\bibinfo {title} {{Observed $D_s(2317)$ and tentative
  $D(2100\text{--}2300)$ as the charmed cousins of the light scalar nonet}},\
  }\href {https://doi.org/10.1103/PhysRevLett.91.012003} {\bibfield  {journal}
  {\bibinfo  {journal} {Phys. Rev. Lett.}\ }\textbf {\bibinfo {volume} {91}},\
  \bibinfo {pages} {012003} (\bibinfo {year} {2003})},\ \Eprint
  {https://arxiv.org/abs/hep-ph/0305035} {arXiv:hep-ph/0305035} \BibitemShut
  {NoStop}%
\bibitem [{\citenamefont {Vijande}\ \emph {et~al.}(2006)\citenamefont
  {Vijande}, \citenamefont {Fernandez},\ and\ \citenamefont
  {Valcarce}}]{Vijande:2006hj}%
  \BibitemOpen
  \bibfield  {author} {\bibinfo {author} {\bibfnamefont {J.}~\bibnamefont
  {Vijande}}, \bibinfo {author} {\bibfnamefont {F.}~\bibnamefont {Fernandez}},\
  and\ \bibinfo {author} {\bibfnamefont {A.}~\bibnamefont {Valcarce}},\
  }\bibfield  {title} {\bibinfo {title} {{Open-charm meson spectroscopy}},\
  }\href {https://doi.org/10.1103/PhysRevD.73.034002} {\bibfield  {journal}
  {\bibinfo  {journal} {Phys. Rev. D}\ }\textbf {\bibinfo {volume} {73}},\
  \bibinfo {pages} {034002} (\bibinfo {year} {2006})},\ \bibinfo {note}
  {[Erratum: Phys.Rev.D 74, 059903 (2006)]},\ \Eprint
  {https://arxiv.org/abs/hep-ph/0601143} {arXiv:hep-ph/0601143} \BibitemShut
  {NoStop}%
\bibitem [{\citenamefont {Eichten}\ \emph {et~al.}(1978)\citenamefont
  {Eichten}, \citenamefont {Gottfried}, \citenamefont {Kinoshita},
  \citenamefont {Lane},\ and\ \citenamefont {Yan}}]{Eichten:1978tg}%
  \BibitemOpen
  \bibfield  {author} {\bibinfo {author} {\bibfnamefont {E.}~\bibnamefont
  {Eichten}}, \bibinfo {author} {\bibfnamefont {K.}~\bibnamefont {Gottfried}},
  \bibinfo {author} {\bibfnamefont {T.}~\bibnamefont {Kinoshita}}, \bibinfo
  {author} {\bibfnamefont {K.~D.}\ \bibnamefont {Lane}},\ and\ \bibinfo
  {author} {\bibfnamefont {T.-M.}\ \bibnamefont {Yan}},\ }\bibfield  {title}
  {\bibinfo {title} {{Charmonium: The Model}},\ }\href
  {https://doi.org/10.1103/PhysRevD.17.3090} {\bibfield  {journal} {\bibinfo
  {journal} {Phys. Rev. D}\ }\textbf {\bibinfo {volume} {17}},\ \bibinfo
  {pages} {3090} (\bibinfo {year} {1978})},\ \bibinfo {note} {[Erratum:
  Phys.Rev.D 21, 313 (1980)]}\BibitemShut {NoStop}%
\bibitem [{\citenamefont {Eichten}\ \emph {et~al.}(1980)\citenamefont
  {Eichten}, \citenamefont {Gottfried}, \citenamefont {Kinoshita},
  \citenamefont {Lane},\ and\ \citenamefont {Yan}}]{Eichten:1979ms}%
  \BibitemOpen
  \bibfield  {author} {\bibinfo {author} {\bibfnamefont {E.}~\bibnamefont
  {Eichten}}, \bibinfo {author} {\bibfnamefont {K.}~\bibnamefont {Gottfried}},
  \bibinfo {author} {\bibfnamefont {T.}~\bibnamefont {Kinoshita}}, \bibinfo
  {author} {\bibfnamefont {K.~D.}\ \bibnamefont {Lane}},\ and\ \bibinfo
  {author} {\bibfnamefont {T.-M.}\ \bibnamefont {Yan}},\ }\bibfield  {title}
  {\bibinfo {title} {{Charmonium: Comparison with Experiment}},\ }\href
  {https://doi.org/10.1103/PhysRevD.21.203} {\bibfield  {journal} {\bibinfo
  {journal} {Phys. Rev. D}\ }\textbf {\bibinfo {volume} {21}},\ \bibinfo
  {pages} {203} (\bibinfo {year} {1980})}\BibitemShut {NoStop}%
\bibitem [{\citenamefont {Gupta}\ \emph {et~al.}(1982)\citenamefont {Gupta},
  \citenamefont {Radford},\ and\ \citenamefont {Repko}}]{Gupta:1982kp}%
  \BibitemOpen
  \bibfield  {author} {\bibinfo {author} {\bibfnamefont {S.~N.}\ \bibnamefont
  {Gupta}}, \bibinfo {author} {\bibfnamefont {S.~F.}\ \bibnamefont {Radford}},\
  and\ \bibinfo {author} {\bibfnamefont {W.~W.}\ \bibnamefont {Repko}},\
  }\bibfield  {title} {\bibinfo {title} {{Quarkonium Spectra and Quantum
  Chromodynamics}},\ }\href {https://doi.org/10.1103/PhysRevD.26.3305}
  {\bibfield  {journal} {\bibinfo  {journal} {Phys. Rev. D}\ }\textbf {\bibinfo
  {volume} {26}},\ \bibinfo {pages} {3305} (\bibinfo {year}
  {1982})}\BibitemShut {NoStop}%
\bibitem [{\citenamefont {Gupta}\ \emph {et~al.}(1983)\citenamefont {Gupta},
  \citenamefont {Radford},\ and\ \citenamefont {Repko}}]{Gupta:1983we}%
  \BibitemOpen
  \bibfield  {author} {\bibinfo {author} {\bibfnamefont {S.~N.}\ \bibnamefont
  {Gupta}}, \bibinfo {author} {\bibfnamefont {S.~F.}\ \bibnamefont {Radford}},\
  and\ \bibinfo {author} {\bibfnamefont {W.~W.}\ \bibnamefont {Repko}},\
  }\bibfield  {title} {\bibinfo {title} {{Quantum Chromodynamic Potential Model
  for Light and Heavy Quarkonia}},\ }\href
  {https://doi.org/10.1103/PhysRevD.28.1716} {\bibfield  {journal} {\bibinfo
  {journal} {Phys. Rev. D}\ }\textbf {\bibinfo {volume} {28}},\ \bibinfo
  {pages} {1716} (\bibinfo {year} {1983})}\BibitemShut {NoStop}%
\bibitem [{\citenamefont {Gupta}\ \emph {et~al.}(1984)\citenamefont {Gupta},
  \citenamefont {Radford},\ and\ \citenamefont {Repko}}]{Gupta:1984jb}%
  \BibitemOpen
  \bibfield  {author} {\bibinfo {author} {\bibfnamefont {S.~N.}\ \bibnamefont
  {Gupta}}, \bibinfo {author} {\bibfnamefont {S.~F.}\ \bibnamefont {Radford}},\
  and\ \bibinfo {author} {\bibfnamefont {W.~W.}\ \bibnamefont {Repko}},\
  }\bibfield  {title} {\bibinfo {title} {{b anti-b SPECTROSCOPY}},\ }\href
  {https://doi.org/10.1103/PhysRevD.30.2424} {\bibfield  {journal} {\bibinfo
  {journal} {Phys. Rev. D}\ }\textbf {\bibinfo {volume} {30}},\ \bibinfo
  {pages} {2424} (\bibinfo {year} {1984})}\BibitemShut {NoStop}%
\bibitem [{\citenamefont {Gupta}\ \emph {et~al.}(1985)\citenamefont {Gupta},
  \citenamefont {Radford},\ and\ \citenamefont {Repko}}]{Gupta:1984um}%
  \BibitemOpen
  \bibfield  {author} {\bibinfo {author} {\bibfnamefont {S.~N.}\ \bibnamefont
  {Gupta}}, \bibinfo {author} {\bibfnamefont {S.~F.}\ \bibnamefont {Radford}},\
  and\ \bibinfo {author} {\bibfnamefont {W.~W.}\ \bibnamefont {Repko}},\
  }\bibfield  {title} {\bibinfo {title} {{Semirelativistic Potential Model for
  Charmonium}},\ }\href {https://doi.org/10.1103/PhysRevD.31.160} {\bibfield
  {journal} {\bibinfo  {journal} {Phys. Rev. D}\ }\textbf {\bibinfo {volume}
  {31}},\ \bibinfo {pages} {160} (\bibinfo {year} {1985})}\BibitemShut
  {NoStop}%
\bibitem [{\citenamefont {Kwong}\ \emph {et~al.}(1988)\citenamefont {Kwong},
  \citenamefont {Mackenzie}, \citenamefont {Rosenfeld},\ and\ \citenamefont
  {Rosner}}]{Kwong:1987ak}%
  \BibitemOpen
  \bibfield  {author} {\bibinfo {author} {\bibfnamefont {W.}~\bibnamefont
  {Kwong}}, \bibinfo {author} {\bibfnamefont {P.~B.}\ \bibnamefont
  {Mackenzie}}, \bibinfo {author} {\bibfnamefont {R.}~\bibnamefont
  {Rosenfeld}},\ and\ \bibinfo {author} {\bibfnamefont {J.~L.}\ \bibnamefont
  {Rosner}},\ }\bibfield  {title} {\bibinfo {title} {{Quarkonium Annihilation
  Rates}},\ }\href {https://doi.org/10.1103/PhysRevD.37.3210} {\bibfield
  {journal} {\bibinfo  {journal} {Phys. Rev. D}\ }\textbf {\bibinfo {volume}
  {37}},\ \bibinfo {pages} {3210} (\bibinfo {year} {1988})}\BibitemShut
  {NoStop}%
\bibitem [{\citenamefont {Kwong}\ and\ \citenamefont
  {Rosner}(1988)}]{Kwong:1988ae}%
  \BibitemOpen
  \bibfield  {author} {\bibinfo {author} {\bibfnamefont {W.}~\bibnamefont
  {Kwong}}\ and\ \bibinfo {author} {\bibfnamefont {J.~L.}\ \bibnamefont
  {Rosner}},\ }\bibfield  {title} {\bibinfo {title} {{$D$ Wave Quarkonium
  Levels of the $\Upsilon$ Family}},\ }\href
  {https://doi.org/10.1103/PhysRevD.38.279} {\bibfield  {journal} {\bibinfo
  {journal} {Phys. Rev. D}\ }\textbf {\bibinfo {volume} {38}},\ \bibinfo
  {pages} {279} (\bibinfo {year} {1988})}\BibitemShut {NoStop}%
\bibitem [{\citenamefont {Vijande}\ \emph {et~al.}(2005)\citenamefont
  {Vijande}, \citenamefont {Fernandez},\ and\ \citenamefont
  {Valcarce}}]{Vijande:2004he}%
  \BibitemOpen
  \bibfield  {author} {\bibinfo {author} {\bibfnamefont {J.}~\bibnamefont
  {Vijande}}, \bibinfo {author} {\bibfnamefont {F.}~\bibnamefont {Fernandez}},\
  and\ \bibinfo {author} {\bibfnamefont {A.}~\bibnamefont {Valcarce}},\
  }\bibfield  {title} {\bibinfo {title} {{Constituent quark model study of the
  meson spectra}},\ }\href {https://doi.org/10.1088/0954-3899/31/5/017}
  {\bibfield  {journal} {\bibinfo  {journal} {J. Phys. G}\ }\textbf {\bibinfo
  {volume} {31}},\ \bibinfo {pages} {481} (\bibinfo {year} {2005})},\ \Eprint
  {https://arxiv.org/abs/hep-ph/0411299} {arXiv:hep-ph/0411299} \BibitemShut
  {NoStop}%
\bibitem [{\citenamefont {Segovia}\ \emph
  {et~al.}(2008{\natexlab{a}})\citenamefont {Segovia}, \citenamefont {Yasser},
  \citenamefont {Entem},\ and\ \citenamefont {Fernandez}}]{Segovia:2008zz}%
  \BibitemOpen
  \bibfield  {author} {\bibinfo {author} {\bibfnamefont {J.}~\bibnamefont
  {Segovia}}, \bibinfo {author} {\bibfnamefont {A.~M.}\ \bibnamefont {Yasser}},
  \bibinfo {author} {\bibfnamefont {D.~R.}\ \bibnamefont {Entem}},\ and\
  \bibinfo {author} {\bibfnamefont {F.}~\bibnamefont {Fernandez}},\ }\bibfield
  {title} {\bibinfo {title} {{JPC=1-- hidden charm resonances}},\ }\href
  {https://doi.org/10.1103/PhysRevD.78.114033} {\bibfield  {journal} {\bibinfo
  {journal} {Phys. Rev. D}\ }\textbf {\bibinfo {volume} {78}},\ \bibinfo
  {pages} {114033} (\bibinfo {year} {2008}{\natexlab{a}})}\BibitemShut
  {NoStop}%
\bibitem [{\citenamefont {Segovia}\ \emph
  {et~al.}(2016{\natexlab{a}})\citenamefont {Segovia}, \citenamefont {Ortega},
  \citenamefont {Entem},\ and\ \citenamefont {Fern\'andez}}]{Segovia:2016xqb}%
  \BibitemOpen
  \bibfield  {author} {\bibinfo {author} {\bibfnamefont {J.}~\bibnamefont
  {Segovia}}, \bibinfo {author} {\bibfnamefont {P.~G.}\ \bibnamefont {Ortega}},
  \bibinfo {author} {\bibfnamefont {D.~R.}\ \bibnamefont {Entem}},\ and\
  \bibinfo {author} {\bibfnamefont {F.}~\bibnamefont {Fern\'andez}},\
  }\bibfield  {title} {\bibinfo {title} {{Bottomonium spectrum revisited}},\
  }\href {https://doi.org/10.1103/PhysRevD.93.074027} {\bibfield  {journal}
  {\bibinfo  {journal} {Phys. Rev. D}\ }\textbf {\bibinfo {volume} {93}},\
  \bibinfo {pages} {074027} (\bibinfo {year} {2016}{\natexlab{a}})},\ \Eprint
  {https://arxiv.org/abs/1601.05093} {arXiv:1601.05093 [hep-ph]} \BibitemShut
  {NoStop}%
\bibitem [{\citenamefont {Segovia}\ \emph
  {et~al.}(2015{\natexlab{a}})\citenamefont {Segovia}, \citenamefont {Entem},\
  and\ \citenamefont {Fernandez}}]{Segovia:2015dia}%
  \BibitemOpen
  \bibfield  {author} {\bibinfo {author} {\bibfnamefont {J.}~\bibnamefont
  {Segovia}}, \bibinfo {author} {\bibfnamefont {D.~R.}\ \bibnamefont {Entem}},\
  and\ \bibinfo {author} {\bibfnamefont {F.}~\bibnamefont {Fernandez}},\
  }\bibfield  {title} {\bibinfo {title} {{Charmed-strange Meson Spectrum: Old
  and New Problems}},\ }\href {https://doi.org/10.1103/PhysRevD.91.094020}
  {\bibfield  {journal} {\bibinfo  {journal} {Phys. Rev. D}\ }\textbf {\bibinfo
  {volume} {91}},\ \bibinfo {pages} {094020} (\bibinfo {year}
  {2015}{\natexlab{a}})},\ \Eprint {https://arxiv.org/abs/1502.03827}
  {arXiv:1502.03827 [hep-ph]} \BibitemShut {NoStop}%
\bibitem [{\citenamefont {Ortega}\ \emph {et~al.}(2020)\citenamefont {Ortega},
  \citenamefont {Segovia}, \citenamefont {Entem},\ and\ \citenamefont
  {Fernandez}}]{Ortega:2020uvc}%
  \BibitemOpen
  \bibfield  {author} {\bibinfo {author} {\bibfnamefont {P.~G.}\ \bibnamefont
  {Ortega}}, \bibinfo {author} {\bibfnamefont {J.}~\bibnamefont {Segovia}},
  \bibinfo {author} {\bibfnamefont {D.~R.}\ \bibnamefont {Entem}},\ and\
  \bibinfo {author} {\bibfnamefont {F.}~\bibnamefont {Fernandez}},\ }\bibfield
  {title} {\bibinfo {title} {{Spectroscopy of $\mathbf {B_c}$ mesons and the
  possibility of finding exotic $\mathbf {B_c}$-like structures}},\ }\href
  {https://doi.org/10.1140/epjc/s10052-020-7764-6} {\bibfield  {journal}
  {\bibinfo  {journal} {Eur. Phys. J. C}\ }\textbf {\bibinfo {volume} {80}},\
  \bibinfo {pages} {223} (\bibinfo {year} {2020})},\ \Eprint
  {https://arxiv.org/abs/2001.08093} {arXiv:2001.08093 [hep-ph]} \BibitemShut
  {NoStop}%
\bibitem [{\citenamefont {Segovia}\ \emph
  {et~al.}(2012{\natexlab{a}})\citenamefont {Segovia}, \citenamefont {Entem},\
  and\ \citenamefont {Fern\'andez}}]{Segovia:2012cd}%
  \BibitemOpen
  \bibfield  {author} {\bibinfo {author} {\bibfnamefont {J.}~\bibnamefont
  {Segovia}}, \bibinfo {author} {\bibfnamefont {D.~R.}\ \bibnamefont {Entem}},\
  and\ \bibinfo {author} {\bibfnamefont {F.}~\bibnamefont {Fern\'andez}},\
  }\bibfield  {title} {\bibinfo {title} {{Scaling of the $^3P_0$ Strength in
  Heavy Meson Strong Decays}},\ }\href
  {https://doi.org/10.1016/j.physletb.2012.08.005} {\bibfield  {journal}
  {\bibinfo  {journal} {Phys. Lett. B}\ }\textbf {\bibinfo {volume} {715}},\
  \bibinfo {pages} {322} (\bibinfo {year} {2012}{\natexlab{a}})},\ \Eprint
  {https://arxiv.org/abs/1205.2215} {arXiv:1205.2215 [hep-ph]} \BibitemShut
  {NoStop}%
\bibitem [{\citenamefont {Segovia}\ \emph {et~al.}(2009)\citenamefont
  {Segovia}, \citenamefont {Yasser}, \citenamefont {Entem},\ and\ \citenamefont
  {Fernandez}}]{Segovia:2009zz}%
  \BibitemOpen
  \bibfield  {author} {\bibinfo {author} {\bibfnamefont {J.}~\bibnamefont
  {Segovia}}, \bibinfo {author} {\bibfnamefont {A.~M.}\ \bibnamefont {Yasser}},
  \bibinfo {author} {\bibfnamefont {D.~R.}\ \bibnamefont {Entem}},\ and\
  \bibinfo {author} {\bibfnamefont {F.}~\bibnamefont {Fernandez}},\ }\bibfield
  {title} {\bibinfo {title} {{Ds-1 (2536) + decays and the properties of P-wave
  charmed strange mesons}},\ }\href
  {https://doi.org/10.1103/PhysRevD.80.054017} {\bibfield  {journal} {\bibinfo
  {journal} {Phys. Rev. D}\ }\textbf {\bibinfo {volume} {80}},\ \bibinfo
  {pages} {054017} (\bibinfo {year} {2009})}\BibitemShut {NoStop}%
\bibitem [{\citenamefont {Segovia}\ \emph
  {et~al.}(2013{\natexlab{a}})\citenamefont {Segovia}, \citenamefont {Entem},\
  and\ \citenamefont {Fernandez}}]{Segovia:2013kg}%
  \BibitemOpen
  \bibfield  {author} {\bibinfo {author} {\bibfnamefont {J.}~\bibnamefont
  {Segovia}}, \bibinfo {author} {\bibfnamefont {D.~R.}\ \bibnamefont {Entem}},\
  and\ \bibinfo {author} {\bibfnamefont {F.}~\bibnamefont {Fernandez}},\
  }\bibfield  {title} {\bibinfo {title} {{Strong charmonium decays in a
  microscopic model}},\ }\href
  {https://doi.org/10.1016/j.nuclphysa.2013.07.004} {\bibfield  {journal}
  {\bibinfo  {journal} {Nucl. Phys. A}\ }\textbf {\bibinfo {volume} {915}},\
  \bibinfo {pages} {125} (\bibinfo {year} {2013}{\natexlab{a}})},\ \Eprint
  {https://arxiv.org/abs/1301.2592} {arXiv:1301.2592 [hep-ph]} \BibitemShut
  {NoStop}%
\bibitem [{\citenamefont {Segovia}\ \emph
  {et~al.}(2015{\natexlab{b}})\citenamefont {Segovia}, \citenamefont {Entem},\
  and\ \citenamefont {Fern\'andez}}]{Segovia:2014mca}%
  \BibitemOpen
  \bibfield  {author} {\bibinfo {author} {\bibfnamefont {J.}~\bibnamefont
  {Segovia}}, \bibinfo {author} {\bibfnamefont {D.~R.}\ \bibnamefont {Entem}},\
  and\ \bibinfo {author} {\bibfnamefont {F.}~\bibnamefont {Fern\'andez}},\
  }\bibfield  {title} {\bibinfo {title} {{Puzzles in hadronic transitions of
  heavy quarkonium with two pion emission}},\ }\href
  {https://doi.org/10.1103/PhysRevD.91.014002} {\bibfield  {journal} {\bibinfo
  {journal} {Phys. Rev. D}\ }\textbf {\bibinfo {volume} {91}},\ \bibinfo
  {pages} {014002} (\bibinfo {year} {2015}{\natexlab{b}})},\ \Eprint
  {https://arxiv.org/abs/1409.7079} {arXiv:1409.7079 [hep-ph]} \BibitemShut
  {NoStop}%
\bibitem [{\citenamefont {Segovia}\ \emph
  {et~al.}(2016{\natexlab{b}})\citenamefont {Segovia}, \citenamefont
  {Fernandez},\ and\ \citenamefont {Entem}}]{Segovia:2015raa}%
  \BibitemOpen
  \bibfield  {author} {\bibinfo {author} {\bibfnamefont {J.}~\bibnamefont
  {Segovia}}, \bibinfo {author} {\bibfnamefont {F.}~\bibnamefont {Fernandez}},\
  and\ \bibinfo {author} {\bibfnamefont {D.~R.}\ \bibnamefont {Entem}},\
  }\bibfield  {title} {\bibinfo {title} {{The Role of Spin-Flipping Terms in
  Hadronic Transitions of ${\Upsilon (4S)}$}},\ }\href
  {https://doi.org/10.1007/s00601-016-1063-7} {\bibfield  {journal} {\bibinfo
  {journal} {Few Body Syst.}\ }\textbf {\bibinfo {volume} {57}},\ \bibinfo
  {pages} {275} (\bibinfo {year} {2016}{\natexlab{b}})},\ \Eprint
  {https://arxiv.org/abs/1507.01607} {arXiv:1507.01607 [hep-ph]} \BibitemShut
  {NoStop}%
\bibitem [{\citenamefont {Mart\'\i{}n-Gonz\'alez}\ \emph
  {et~al.}(2022)\citenamefont {Mart\'\i{}n-Gonz\'alez}, \citenamefont {Ortega},
  \citenamefont {Entem}, \citenamefont {Fern\'andez},\ and\ \citenamefont
  {Segovia}}]{Martin-Gonzalez:2022qwd}%
  \BibitemOpen
  \bibfield  {author} {\bibinfo {author} {\bibfnamefont {B.}~\bibnamefont
  {Mart\'\i{}n-Gonz\'alez}}, \bibinfo {author} {\bibfnamefont {P.~G.}\
  \bibnamefont {Ortega}}, \bibinfo {author} {\bibfnamefont {D.~R.}\
  \bibnamefont {Entem}}, \bibinfo {author} {\bibfnamefont {F.}~\bibnamefont
  {Fern\'andez}},\ and\ \bibinfo {author} {\bibfnamefont {J.}~\bibnamefont
  {Segovia}},\ }\bibfield  {title} {\bibinfo {title} {{Toward the discovery of
  novel Bc states: Radiative and hadronic transitions}},\ }\href
  {https://doi.org/10.1103/PhysRevD.106.054009} {\bibfield  {journal} {\bibinfo
   {journal} {Phys. Rev. D}\ }\textbf {\bibinfo {volume} {106}},\ \bibinfo
  {pages} {054009} (\bibinfo {year} {2022})},\ \Eprint
  {https://arxiv.org/abs/2205.05950} {arXiv:2205.05950 [hep-ph]} \BibitemShut
  {NoStop}%
\bibitem [{\citenamefont {Segovia}\ \emph
  {et~al.}(2011{\natexlab{a}})\citenamefont {Segovia}, \citenamefont {Entem},\
  and\ \citenamefont {Fernandez}}]{Segovia:2011zza}%
  \BibitemOpen
  \bibfield  {author} {\bibinfo {author} {\bibfnamefont {J.}~\bibnamefont
  {Segovia}}, \bibinfo {author} {\bibfnamefont {D.~R.}\ \bibnamefont {Entem}},\
  and\ \bibinfo {author} {\bibfnamefont {F.}~\bibnamefont {Fernandez}},\
  }\bibfield  {title} {\bibinfo {title} {{Charmonium resonances in e+ e-
  exclusive reactions around the psi(4415) region}},\ }\href
  {https://doi.org/10.1103/PhysRevD.83.114018} {\bibfield  {journal} {\bibinfo
  {journal} {Phys. Rev. D}\ }\textbf {\bibinfo {volume} {83}},\ \bibinfo
  {pages} {114018} (\bibinfo {year} {2011}{\natexlab{a}})}\BibitemShut
  {NoStop}%
\bibitem [{\citenamefont {Segovia}\ \emph
  {et~al.}(2011{\natexlab{b}})\citenamefont {Segovia}, \citenamefont
  {Albertus}, \citenamefont {Entem}, \citenamefont {Fernandez}, \citenamefont
  {Hernandez},\ and\ \citenamefont {Perez-Garcia}}]{Segovia:2011dg}%
  \BibitemOpen
  \bibfield  {author} {\bibinfo {author} {\bibfnamefont {J.}~\bibnamefont
  {Segovia}}, \bibinfo {author} {\bibfnamefont {C.}~\bibnamefont {Albertus}},
  \bibinfo {author} {\bibfnamefont {D.~R.}\ \bibnamefont {Entem}}, \bibinfo
  {author} {\bibfnamefont {F.}~\bibnamefont {Fernandez}}, \bibinfo {author}
  {\bibfnamefont {E.}~\bibnamefont {Hernandez}},\ and\ \bibinfo {author}
  {\bibfnamefont {M.~A.}\ \bibnamefont {Perez-Garcia}},\ }\bibfield  {title}
  {\bibinfo {title} {{Semileptonic $B$ and $B_{s}$ decays into orbitally
  excited charmed mesons}},\ }\href
  {https://doi.org/10.1103/PhysRevD.84.094029} {\bibfield  {journal} {\bibinfo
  {journal} {Phys. Rev. D}\ }\textbf {\bibinfo {volume} {84}},\ \bibinfo
  {pages} {094029} (\bibinfo {year} {2011}{\natexlab{b}})},\ \Eprint
  {https://arxiv.org/abs/1107.4248} {arXiv:1107.4248 [hep-ph]} \BibitemShut
  {NoStop}%
\bibitem [{\citenamefont {Segovia}\ \emph
  {et~al.}(2012{\natexlab{b}})\citenamefont {Segovia}, \citenamefont
  {Albertus}, \citenamefont {Hernandez}, \citenamefont {Fernandez},\ and\
  \citenamefont {Entem}}]{Segovia:2012yh}%
  \BibitemOpen
  \bibfield  {author} {\bibinfo {author} {\bibfnamefont {J.}~\bibnamefont
  {Segovia}}, \bibinfo {author} {\bibfnamefont {C.}~\bibnamefont {Albertus}},
  \bibinfo {author} {\bibfnamefont {E.}~\bibnamefont {Hernandez}}, \bibinfo
  {author} {\bibfnamefont {F.}~\bibnamefont {Fernandez}},\ and\ \bibinfo
  {author} {\bibfnamefont {D.~R.}\ \bibnamefont {Entem}},\ }\bibfield  {title}
  {\bibinfo {title} {{Nonleptonic $B \to D^{(*)}D_{sJ}^{(*)}$ decays and the
  nature of the orbitally excited charmed-strange mesons}},\ }\href
  {https://doi.org/10.1103/PhysRevD.86.014010} {\bibfield  {journal} {\bibinfo
  {journal} {Phys. Rev. D}\ }\textbf {\bibinfo {volume} {86}},\ \bibinfo
  {pages} {014010} (\bibinfo {year} {2012}{\natexlab{b}})},\ \Eprint
  {https://arxiv.org/abs/1203.4362} {arXiv:1203.4362 [hep-ph]} \BibitemShut
  {NoStop}%
\bibitem [{\citenamefont {Lakhina}\ and\ \citenamefont
  {Swanson}(2007)}]{Lakhina:2006fy}%
  \BibitemOpen
  \bibfield  {author} {\bibinfo {author} {\bibfnamefont {O.}~\bibnamefont
  {Lakhina}}\ and\ \bibinfo {author} {\bibfnamefont {E.~S.}\ \bibnamefont
  {Swanson}},\ }\bibfield  {title} {\bibinfo {title} {{A Canonical
  Ds(2317)?}},\ }\href {https://doi.org/10.1016/j.physletb.2007.01.075}
  {\bibfield  {journal} {\bibinfo  {journal} {Phys. Lett. B}\ }\textbf
  {\bibinfo {volume} {650}},\ \bibinfo {pages} {159} (\bibinfo {year}
  {2007})},\ \Eprint {https://arxiv.org/abs/hep-ph/0608011}
  {arXiv:hep-ph/0608011} \BibitemShut {NoStop}%
\bibitem [{\citenamefont {Gupta}\ and\ \citenamefont
  {Radford}(1981)}]{Gupta:1981pd}%
  \BibitemOpen
  \bibfield  {author} {\bibinfo {author} {\bibfnamefont {S.~N.}\ \bibnamefont
  {Gupta}}\ and\ \bibinfo {author} {\bibfnamefont {S.~F.}\ \bibnamefont
  {Radford}},\ }\bibfield  {title} {\bibinfo {title} {{Quark Quark and Quark -
  Anti-quark Potentials}},\ }\href {https://doi.org/10.1103/PhysRevD.24.2309}
  {\bibfield  {journal} {\bibinfo  {journal} {Phys. Rev. D}\ }\textbf {\bibinfo
  {volume} {24}},\ \bibinfo {pages} {2309} (\bibinfo {year}
  {1981})}\BibitemShut {NoStop}%
\bibitem [{\citenamefont {Segovia}\ \emph
  {et~al.}(2013{\natexlab{b}})\citenamefont {Segovia}, \citenamefont {Entem},
  \citenamefont {Fernandez},\ and\ \citenamefont
  {Hernandez}}]{Segovia:2013wma}%
  \BibitemOpen
  \bibfield  {author} {\bibinfo {author} {\bibfnamefont {J.}~\bibnamefont
  {Segovia}}, \bibinfo {author} {\bibfnamefont {D.~R.}\ \bibnamefont {Entem}},
  \bibinfo {author} {\bibfnamefont {F.}~\bibnamefont {Fernandez}},\ and\
  \bibinfo {author} {\bibfnamefont {E.}~\bibnamefont {Hernandez}},\ }\bibfield
  {title} {\bibinfo {title} {{Constituent quark model description of charmonium
  phenomenology}},\ }\href {https://doi.org/10.1142/S0218301313300269}
  {\bibfield  {journal} {\bibinfo  {journal} {Int. J. Mod. Phys. E}\ }\textbf
  {\bibinfo {volume} {22}},\ \bibinfo {pages} {1330026} (\bibinfo {year}
  {2013}{\natexlab{b}})},\ \Eprint {https://arxiv.org/abs/1309.6926}
  {arXiv:1309.6926 [hep-ph]} \BibitemShut {NoStop}%
\bibitem [{\citenamefont {Diakonov}(2003)}]{Diakonov:2002fq}%
  \BibitemOpen
  \bibfield  {author} {\bibinfo {author} {\bibfnamefont {D.}~\bibnamefont
  {Diakonov}},\ }\bibfield  {title} {\bibinfo {title} {{Instantons at work}},\
  }\href {https://doi.org/10.1016/S0146-6410(03)90014-7} {\bibfield  {journal}
  {\bibinfo  {journal} {Prog. Part. Nucl. Phys.}\ }\textbf {\bibinfo {volume}
  {51}},\ \bibinfo {pages} {173} (\bibinfo {year} {2003})},\ \Eprint
  {https://arxiv.org/abs/hep-ph/0212026} {arXiv:hep-ph/0212026} \BibitemShut
  {NoStop}%
\bibitem [{\citenamefont {Segovia}\ \emph
  {et~al.}(2008{\natexlab{b}})\citenamefont {Segovia}, \citenamefont {Entem},\
  and\ \citenamefont {Fernandez}}]{Segovia:2008zza}%
  \BibitemOpen
  \bibfield  {author} {\bibinfo {author} {\bibfnamefont {J.}~\bibnamefont
  {Segovia}}, \bibinfo {author} {\bibfnamefont {D.~R.}\ \bibnamefont {Entem}},\
  and\ \bibinfo {author} {\bibfnamefont {F.}~\bibnamefont {Fernandez}},\
  }\bibfield  {title} {\bibinfo {title} {{Is chiral symmetry restored in the
  excited meson spectrum?}},\ }\href
  {https://doi.org/10.1016/j.physletb.2008.02.051} {\bibfield  {journal}
  {\bibinfo  {journal} {Phys. Lett. B}\ }\textbf {\bibinfo {volume} {662}},\
  \bibinfo {pages} {33} (\bibinfo {year} {2008}{\natexlab{b}})}\BibitemShut
  {NoStop}%
\bibitem [{\citenamefont {Ortega}\ \emph {et~al.}(2016)\citenamefont {Ortega},
  \citenamefont {Segovia}, \citenamefont {Entem},\ and\ \citenamefont
  {Fernandez}}]{Ortega:2016mms}%
  \BibitemOpen
  \bibfield  {author} {\bibinfo {author} {\bibfnamefont {P.~G.}\ \bibnamefont
  {Ortega}}, \bibinfo {author} {\bibfnamefont {J.}~\bibnamefont {Segovia}},
  \bibinfo {author} {\bibfnamefont {D.~R.}\ \bibnamefont {Entem}},\ and\
  \bibinfo {author} {\bibfnamefont {F.}~\bibnamefont {Fernandez}},\ }\bibfield
  {title} {\bibinfo {title} {{Molecular components in P-wave charmed-strange
  mesons}},\ }\href {https://doi.org/10.1103/PhysRevD.94.074037} {\bibfield
  {journal} {\bibinfo  {journal} {Phys. Rev. D}\ }\textbf {\bibinfo {volume}
  {94}},\ \bibinfo {pages} {074037} (\bibinfo {year} {2016})},\ \Eprint
  {https://arxiv.org/abs/1603.07000} {arXiv:1603.07000 [hep-ph]} \BibitemShut
  {NoStop}%
\bibitem [{\citenamefont {Bali}\ \emph {et~al.}(2005)\citenamefont {Bali},
  \citenamefont {Neff}, \citenamefont {Duessel}, \citenamefont {Lippert},\ and\
  \citenamefont {Schilling}}]{Bali:2005fu}%
  \BibitemOpen
  \bibfield  {author} {\bibinfo {author} {\bibfnamefont {G.~S.}\ \bibnamefont
  {Bali}}, \bibinfo {author} {\bibfnamefont {H.}~\bibnamefont {Neff}}, \bibinfo
  {author} {\bibfnamefont {T.}~\bibnamefont {Duessel}}, \bibinfo {author}
  {\bibfnamefont {T.}~\bibnamefont {Lippert}},\ and\ \bibinfo {author}
  {\bibfnamefont {K.}~\bibnamefont {Schilling}} (\bibinfo {collaboration}
  {SESAM}),\ }\bibfield  {title} {\bibinfo {title} {{Observation of string
  breaking in QCD}},\ }\href {https://doi.org/10.1103/PhysRevD.71.114513}
  {\bibfield  {journal} {\bibinfo  {journal} {Phys. Rev. D}\ }\textbf {\bibinfo
  {volume} {71}},\ \bibinfo {pages} {114513} (\bibinfo {year} {2005})},\
  \Eprint {https://arxiv.org/abs/hep-lat/0505012} {arXiv:hep-lat/0505012}
  \BibitemShut {NoStop}%
\bibitem [{\citenamefont {Hiyama}\ \emph {et~al.}(2003)\citenamefont {Hiyama},
  \citenamefont {Kino},\ and\ \citenamefont {Kamimura}}]{Hiyama:2003cu}%
  \BibitemOpen
  \bibfield  {author} {\bibinfo {author} {\bibfnamefont {E.}~\bibnamefont
  {Hiyama}}, \bibinfo {author} {\bibfnamefont {Y.}~\bibnamefont {Kino}},\ and\
  \bibinfo {author} {\bibfnamefont {M.}~\bibnamefont {Kamimura}},\ }\bibfield
  {title} {\bibinfo {title} {{Gaussian expansion method for few-body
  systems}},\ }\href {https://doi.org/10.1016/S0146-6410(03)90015-9} {\bibfield
   {journal} {\bibinfo  {journal} {Prog. Part. Nucl. Phys.}\ }\textbf {\bibinfo
  {volume} {51}},\ \bibinfo {pages} {223} (\bibinfo {year} {2003})}\BibitemShut
  {NoStop}%
\bibitem [{\citenamefont {Capstick}\ and\ \citenamefont
  {Isgur}(1985)}]{Capstick:1985xss}%
  \BibitemOpen
  \bibfield  {author} {\bibinfo {author} {\bibfnamefont {S.}~\bibnamefont
  {Capstick}}\ and\ \bibinfo {author} {\bibfnamefont {N.}~\bibnamefont
  {Isgur}},\ }\bibfield  {title} {\bibinfo {title} {{Baryons in a Relativized
  Quark Model with Chromodynamics}},\ }\href {https://doi.org/10.1063/1.35361}
  {\bibfield  {journal} {\bibinfo  {journal} {AIP Conf. Proc.}\ }\textbf
  {\bibinfo {volume} {132}},\ \bibinfo {pages} {267} (\bibinfo {year}
  {1985})}\BibitemShut {NoStop}%
\bibitem [{\citenamefont {Manohar}\ and\ \citenamefont
  {Georgi}(1984)}]{Manohar:1983md}%
  \BibitemOpen
  \bibfield  {author} {\bibinfo {author} {\bibfnamefont {A.}~\bibnamefont
  {Manohar}}\ and\ \bibinfo {author} {\bibfnamefont {H.}~\bibnamefont
  {Georgi}},\ }\bibfield  {title} {\bibinfo {title} {{Chiral Quarks and the
  Nonrelativistic Quark Model}},\ }\href
  {https://doi.org/10.1016/0550-3213(84)90231-1} {\bibfield  {journal}
  {\bibinfo  {journal} {Nucl. Phys. B}\ }\textbf {\bibinfo {volume} {234}},\
  \bibinfo {pages} {189} (\bibinfo {year} {1984})}\BibitemShut {NoStop}%
\bibitem [{\citenamefont {Lucha}\ \emph {et~al.}(1991)\citenamefont {Lucha},
  \citenamefont {Schoberl},\ and\ \citenamefont {Gromes}}]{Lucha:1991vn}%
  \BibitemOpen
  \bibfield  {author} {\bibinfo {author} {\bibfnamefont {W.}~\bibnamefont
  {Lucha}}, \bibinfo {author} {\bibfnamefont {F.~F.}\ \bibnamefont
  {Schoberl}},\ and\ \bibinfo {author} {\bibfnamefont {D.}~\bibnamefont
  {Gromes}},\ }\bibfield  {title} {\bibinfo {title} {{Bound states of
  quarks}},\ }\href {https://doi.org/10.1016/0370-1573(91)90001-3} {\bibfield
  {journal} {\bibinfo  {journal} {Phys. Rept.}\ }\textbf {\bibinfo {volume}
  {200}},\ \bibinfo {pages} {127} (\bibinfo {year} {1991})}\BibitemShut
  {NoStop}%
\bibitem [{\citenamefont {Ortega}\ \emph
  {et~al.}(2023{\natexlab{a}})\citenamefont {Ortega}, \citenamefont {Segovia},
  \citenamefont {Entem},\ and\ \citenamefont {Fernandez}}]{Ortega:2022efc}%
  \BibitemOpen
  \bibfield  {author} {\bibinfo {author} {\bibfnamefont {P.~G.}\ \bibnamefont
  {Ortega}}, \bibinfo {author} {\bibfnamefont {J.}~\bibnamefont {Segovia}},
  \bibinfo {author} {\bibfnamefont {D.~R.}\ \bibnamefont {Entem}},\ and\
  \bibinfo {author} {\bibfnamefont {F.}~\bibnamefont {Fernandez}},\ }\bibfield
  {title} {\bibinfo {title} {{Nature of the doubly-charmed tetraquark Tcc+ in a
  constituent quark model}},\ }\href
  {https://doi.org/10.1016/j.physletb.2023.137918} {\bibfield  {journal}
  {\bibinfo  {journal} {Phys. Lett. B}\ }\textbf {\bibinfo {volume} {841}},\
  \bibinfo {pages} {137918} (\bibinfo {year} {2023}{\natexlab{a}})},\ \bibinfo
  {note} {[Erratum: Phys.Lett.B 847, 138308 (2023)]},\ \Eprint
  {https://arxiv.org/abs/2211.06118} {arXiv:2211.06118 [hep-ph]} \BibitemShut
  {NoStop}%
\bibitem [{\citenamefont {Ortega}\ \emph
  {et~al.}(2023{\natexlab{b}})\citenamefont {Ortega}, \citenamefont {Entem},
  \citenamefont {Fernandez},\ and\ \citenamefont {Segovia}}]{Ortega:2023azl}%
  \BibitemOpen
  \bibfield  {author} {\bibinfo {author} {\bibfnamefont {P.~G.}\ \bibnamefont
  {Ortega}}, \bibinfo {author} {\bibfnamefont {D.~R.}\ \bibnamefont {Entem}},
  \bibinfo {author} {\bibfnamefont {F.}~\bibnamefont {Fernandez}},\ and\
  \bibinfo {author} {\bibfnamefont {J.}~\bibnamefont {Segovia}},\ }\bibfield
  {title} {\bibinfo {title} {{Novel Tcs and Tcs\textasciimacron{} candidates in
  a constituent-quark-model-based meson-meson coupled-channels calculation}},\
  }\href {https://doi.org/10.1103/PhysRevD.108.094035} {\bibfield  {journal}
  {\bibinfo  {journal} {Phys. Rev. D}\ }\textbf {\bibinfo {volume} {108}},\
  \bibinfo {pages} {094035} (\bibinfo {year} {2023}{\natexlab{b}})},\ \Eprint
  {https://arxiv.org/abs/2305.14430} {arXiv:2305.14430 [hep-ph]} \BibitemShut
  {NoStop}%
\bibitem [{\citenamefont {Ortega}\ \emph {et~al.}(2019)\citenamefont {Ortega},
  \citenamefont {Segovia}, \citenamefont {Entem},\ and\ \citenamefont
  {Fern\'andez}}]{Ortega:2018cnm}%
  \BibitemOpen
  \bibfield  {author} {\bibinfo {author} {\bibfnamefont {P.~G.}\ \bibnamefont
  {Ortega}}, \bibinfo {author} {\bibfnamefont {J.}~\bibnamefont {Segovia}},
  \bibinfo {author} {\bibfnamefont {D.~R.}\ \bibnamefont {Entem}},\ and\
  \bibinfo {author} {\bibfnamefont {F.}~\bibnamefont {Fern\'andez}},\
  }\bibfield  {title} {\bibinfo {title} {{The $Z_c$ structures in a
  coupled-channels model}},\ }\href
  {https://doi.org/10.1140/epjc/s10052-019-6552-7} {\bibfield  {journal}
  {\bibinfo  {journal} {Eur. Phys. J. C}\ }\textbf {\bibinfo {volume} {79}},\
  \bibinfo {pages} {78} (\bibinfo {year} {2019})},\ \Eprint
  {https://arxiv.org/abs/1808.00914} {arXiv:1808.00914 [hep-ph]} \BibitemShut
  {NoStop}%
\bibitem [{\citenamefont {Ortega}\ \emph {et~al.}(2021)\citenamefont {Ortega},
  \citenamefont {Segovia},\ and\ \citenamefont {Fernandez}}]{Ortega:2021xst}%
  \BibitemOpen
  \bibfield  {author} {\bibinfo {author} {\bibfnamefont {P.~G.}\ \bibnamefont
  {Ortega}}, \bibinfo {author} {\bibfnamefont {J.}~\bibnamefont {Segovia}},\
  and\ \bibinfo {author} {\bibfnamefont {F.}~\bibnamefont {Fernandez}},\
  }\bibfield  {title} {\bibinfo {title} {{Zb structures in a constituent quark
  model coupled-channels calculation}},\ }\href
  {https://doi.org/10.1103/PhysRevD.104.094004} {\bibfield  {journal} {\bibinfo
   {journal} {Phys. Rev. D}\ }\textbf {\bibinfo {volume} {104}},\ \bibinfo
  {pages} {094004} (\bibinfo {year} {2021})},\ \Eprint
  {https://arxiv.org/abs/2107.02544} {arXiv:2107.02544 [hep-ph]} \BibitemShut
  {NoStop}%
\bibitem [{\citenamefont {Ortega}\ \emph {et~al.}(2010)\citenamefont {Ortega},
  \citenamefont {Segovia}, \citenamefont {Entem},\ and\ \citenamefont
  {Fernandez}}]{Ortega:2009hj}%
  \BibitemOpen
  \bibfield  {author} {\bibinfo {author} {\bibfnamefont {P.~G.}\ \bibnamefont
  {Ortega}}, \bibinfo {author} {\bibfnamefont {J.}~\bibnamefont {Segovia}},
  \bibinfo {author} {\bibfnamefont {D.~R.}\ \bibnamefont {Entem}},\ and\
  \bibinfo {author} {\bibfnamefont {F.}~\bibnamefont {Fernandez}},\ }\bibfield
  {title} {\bibinfo {title} {{Coupled channel approach to the structure of the
  X(3872)}},\ }\href {https://doi.org/10.1103/PhysRevD.81.054023} {\bibfield
  {journal} {\bibinfo  {journal} {Phys. Rev. D}\ }\textbf {\bibinfo {volume}
  {81}},\ \bibinfo {pages} {054023} (\bibinfo {year} {2010})},\ \Eprint
  {https://arxiv.org/abs/0907.3997} {arXiv:0907.3997 [hep-ph]} \BibitemShut
  {NoStop}%
\end{thebibliography}%

\end{document}